\documentclass{aa}  
\usepackage{natbib}
\bibpunct{(}{)}{;}{a}{}{,} 
\usepackage{graphicx}
\usepackage{txfonts}
\usepackage{amsmath}
\usepackage{appendix}
\usepackage{subfigure}
\usepackage{color}
\usepackage{hyperref}
\usepackage{multirow}
\usepackage[switch]{lineno}

\hypersetup{colorlinks,allcolors={blue}}

\graphicspath{{./}{figure/}}

\begin{document}

   \title{Numerical Investigation of Instabilities in Over-pressured Magnetized Relativistic Jets}

   \author{Xu-Fan Hu
          \inst{1}
          \and
          Yosuke Mizuno\inst{1,2,3}
          \and
          Christian M. Fromm \inst{4,3,5}
          }

   \institute{Tsung-Dao Lee Institute, Shanghai Jiao-Tong University, Shanghai, 1 Lisuo Road, 201210, People's Republic of China\\
              \email{ebr105@sjtu.edu.cn; mizuno@sjtu.edu.cn}
        \and
            School of Physics \& Astronomy, Shanghai Jiao-Tong University, Shanghai, 800 Dongchuan Road, 200240, People's Republic of China
        \and
            Institut f\"ur Theoretische Physik, Goethe Universit\"at, Max-von-Laue-Str. 1, D-60438 Frankfurt, Germany \label{affil:frankfurt}
        \and
            Institut f\"ur Theoretische Physik und Astrophysik, Universit\"at W\"urzburg, Emil-Fischer-Str. 31, D-97074 W\"urzburg, Germany\label{affil:wuerzburg}
        \and
            Max-Planck-Institut f\"ur Radioastronomie, Auf dem H\"ugel 69, D-53121 Bonn, Germany
        }

   \date{Received July 15, 2024; accepted September 16, 2024}

 
  \abstract
   {Relativistic jets from Active Galactic Nuclei are observed to be collimated on the parsec scale. When the pressure between the jet and the ambient medium is mismatched, recollimation shocks and rarefaction shocks are formed. Previous numerical simulations have shown that instabilities can destroy the recollimation structure of jets.}
   {In this study, we aim to study the instabilities of non-equilibrium over-pressured relativistic jets with helical magnetic fields. Especially, we investigate how the magnetic pitch affects the development of instabilities.}
   {We perform three-dimensional relativistic magnetohydrodynamic simulations for different magnetic pitches, as well as a two-dimension simulation and a relativistic hydrodynamic simulation served as comparison groups}
   {In our simulations, Rayleigh-Taylor Instability (RTI) is triggered at the interface between the jet and ambient medium in the recollimation structure of the jet.
    We found that when the magnetic pitch decreases the growth of RTI becomes weak but interestingly, another instability, the CD kink instability is excited.
    The excitement of CD kink instability after passing the recollimation shocks can match the explanation of the quasi-periodic oscillations observed in BL Lac qualitatively.}
   {}


   \maketitle
%

\section{Introduction}

Relativistic jets, one of the most remarkable features of Supermassive Black Holes (SMBHs), have drawn the wide attention of scientists since it was first recorded in 1918 \citep{curtis1918}. Relativistic jets have enormous power and move at relativistic speeds (i.e., high Lorentz factors), causing the observations of superluminal motion. Although a hundred years passed since the first discovery, the mechanism of jet launching is still under debate. 
There are two major mechanisms for jet launching: one is the direct energy extraction of a rotating black hole by the Blandford-Znajek mechanism \citep{blandford1977} and the other is the acceleration by magneto-centrifugal force from the accretion disk that is known as Blandford-Payne mechanism \citep{blandford1982}.


Through the observations from the radio to the X-ray regime, we know relativistic jets in AGNs are highly collimated on a scale from sub-pc to Mpc \citep{pushkarev2009}. 
When a jet propagates through an ambient medium, a pressure mismatch between the jet and the ambient medium will arise as a result of changing the pressure of the ambient medium. The pressure mismatch drives a radial oscillating motion of the jet that leads to multiple recollimation regions inside the jet \citep[e.g.,][]{gomez1997, komissarov1997,agudo2001,porth2015,mizuno2015,fromm2016}. Such recollimation shocks would be connected to the quasi-stationary features seen in jets of Active Galactic Nuclei (AGNs) \citep[e.g.,][]{jorstad2005,lister2013,cohen2014,gomez2016}.
The case of M87 is very interesting because the stationary feature HST-1 knot is thought to be related to a recollimation shock and we see the emergence of new superluminal components from HST-1 \citep[][]{giroletti2012}.  In addition, very high-energy emission is associated with the variability of HST-1 region \citep[][]{cheung2007}.


During the propagation of the jets to large scales several instabilities can grow which could lead to a loss of the collimation and even destroy the jet. 
The observed morphological differences of relativistic jets could also be linked to a physical reason for the Fanaroff-Riley dichotomy \citep{fanaroff1974}. 
For example, a Poynting-flux dominated jet with a large-scale helical magnetic field is unstable under the current-driven instability (CDI), in particular, to the kink mode \citep[e.g.,][]{moll2008,mizuno2009,mizuno2014}. In such a situation, the shape of the jet changes to a helically twisted structure. Another possible instability is Kelvin-Helmholtz instability (KHI). In the presence of velocity shear, it leads to small amplitude wave-like disturbances on the interface. Centrifugal instability (CFI) can be excited within the rotational jet \citep{Gourgouliatos2018} that forms mushroom shapes on the interface at a regular angle interval in 2D sections perpendicular to the jet propagation direction. However, a weak toroidal magnetic field stabilizes against this instability \citep{Komissarov2019,matsumoto2021}.
The Rayleigh-Taylor instability (RTI) is also possible to grow at the boundary of the jet and external medium which is characterized by a finger-like structure \citep{bottcher2012}. 
Generally, it is difficult to derive the stability condition for each instability in the jet analytically because the jets are subject to a complicated non-linear interplay between jet rotation, magnetic fields, relativistic effects, the structure between the jet and external medium, etc.

Adopted by a large number of previous works, relativistic magnetohydrodynamic (RMHD) simulations are proven to be an effective method for studying the propagation and stability of the jet. 
\cite{mizuno2015} have simulated over-pressured jets with and without magnetic fields in 2D. All of the jets showed the development of recollimation shock structures and no instability was triggered. 
\cite{Matsumoto2013} have performed 2D relativistic hydrodynamic (RHD) simulations of a co-moving over-pressured jet on the plane perpendicular to the jet. They found when the effective inertia $\gamma^2\rho h$ of the jet is greater than that of the outer external medium, the boundary between the jet and external medium is unstable against RTI, where $\gamma$ is Lorentz factor, $\rho$ is density, and $h$ is specific enthalpy.
The effects of the jet rotation and toroidal magnetic field for the RTI in two-component jets have been investigated by \cite{millas2017} and shown that magnetic field provides stability against RTI.
\cite{abolmasov2023} pointed out that three vectors in the RTI cannot be co-planar: jet propagation direction, the effective gravity direction, and the unstable mode's wave vector direction. Thus, three-dimensional, non-axisymmetric simulations are essential for the study of RTI.\cite{Gourgouliatos2018} simulated 3D RHD jets and showed that the instabilities in recollimation shocks are rather related to the centrifugal instability (CFI) than to the RTI. The studies are extended to 3D MHD simulations by \cite{matsumoto2021}. Recently,  \cite{costa2023} tried to use 3D HD simulations to explain morphologies of FR0 type of jets \citep{ghisellini2011} observed by VLBI.

In this paper, we extend our previous 2D RMHD simulations of over-pressured jets with the helical magnetic field by \cite{mizuno2015} to 3D non-axisymmetric simulations. As we discussed, three-dimensional, non-axisymmetric simulations are essential for the study of instabilities in the jet.
We investigate how the helical magnetic field affects the stability of overpressured jets.

The structure of this paper is as follows. In Section \ref{num}, we present our method and the setup of the 3D RMHD simulations. In Section \ref{2d} and Section \ref{3dhd} we first show the 2D RMHD and 3D RHD simulations of an over-pressured jet. In Section \ref{3d}, we compare the results of the 3D RMHD simulations with helical magnetic field and discuss the effect of magnetic pitches. We analyze the recollimation shock structure and the possibility of the excitement of kink instability in Section \ref{dis} and summarize and discuss our findings in Section \ref{dis}. 

\section{Numerical setup}\label{num}
\begin{table*}[!ht]
    \centering
    \begin{tabular}{ll}
    \hline \hline
        Frame & 3D Cartesian \\
        EOS & ideal gas \\ \hline
        \multirow{3}{*}{Riemann Solver} & MHD1-2D: HLL\\
                                        & HD: HLLC\\
                                        & Others: TVDLF\\ \hline
        Divergence control & Constrained Transport (MHD3: Eight-Waves) \\
        Reconstruction & piecewise parabolic method\\
        Time stepping & 3rd-order TVD Runge Kutta\\
        Initial state & a preexisting cylindrical jet \\
        Density [$\rho_0$]& $\rho_a=$1, $\rho_j=$0.01\\
        Velocity [c]& $v_j=0.9428$  $\gamma_j=3$,  $v_a=0$\\
        Pressure [$\rho_0 c^2$]& $p_j=1.5p_a=1.5p_0$,  $p_0=0.018$\\
        Domain & $(x,y,z)=(\pm5R_j,\pm5R_j,60R_j)$\\
        Number of cells & $(N_x,N_y,N_z)=(500,500,600)$\\
        Boundary conditions(in,out) & (outflow,outflow)$\times$(outflow,outflow)$\times$(inflow,outflow) \\ 
        time[$R_j/c$]&400\\ \hline \hline
    \end{tabular}
    \caption{Setup of our 3D RMHD simulations}
    \label{T1}
\end{table*}

We perform 3D relativistic magnetohydrodynamic (RMHD) simulations of over-pressured jets
using public RMHD code {\tt PLUTO} \citep{mignone2007} adopting cartesian coordinates. We solve the following form of the RMHD equations:
\begin{align}
\frac{\partial}{\partial t}(\gamma\rho)+\nabla\cdot(\gamma\rho \mathbf{v})=0, \label{1} \\   
\frac{\partial}{\partial t}(\omega_t\gamma^2\mathbf{v}-b^0\mathbf{b})+\nabla\cdot(\omega_t\gamma^2\mathbf{v}\mathbf{v}-\mathbf{b}\mathbf{b}+\mathbf{I} p_t)=0, \label{2} \\
\frac{\partial}{\partial t}(\omega_t\gamma^2-b^0b^0-p_t)+\nabla\cdot(\omega_t\gamma^2\mathbf{v}-b^0\mathbf{b})=0, \label{3} \\    
\frac{\partial}{\partial t}\mathbf{B}+\nabla\cdot(\mathbf{v}\mathbf{B}-\mathbf{B}\mathbf{v})=0, \label{4}
\end{align}
where $b^0=\gamma\mathbf{v\cdot B}$ , $\mathbf{b}=\mathbf{B}/\gamma+\gamma\mathbf{(v\cdot B)v}$, $\omega_t=\rho h+B^2/\gamma+\mathbf{(v\cdot B)^2}$, and $p_t=p_{gas}+[B^2/\gamma^2+\mathbf{(v\cdot B)^2}]/2$. Here we set $c=1$ and the units of velocity, density, pressure, and magnetic field strength are $c$, $\rho_0$, $\rho_0c^2$, and $\sqrt{4\pi\rho_0c^2}$, respectively.  The specific enthalpy is defined as $h=1+\Gamma p_{gas} /(\Gamma-1)\rho$ and magnetization is $\sigma=|\mathbf{b}|^2/\rho$. We use the ideal equation of state with the adiabatic index $\Gamma=4/3$.

\begin{figure}
    \centering
    \includegraphics[width=0.8\columnwidth]{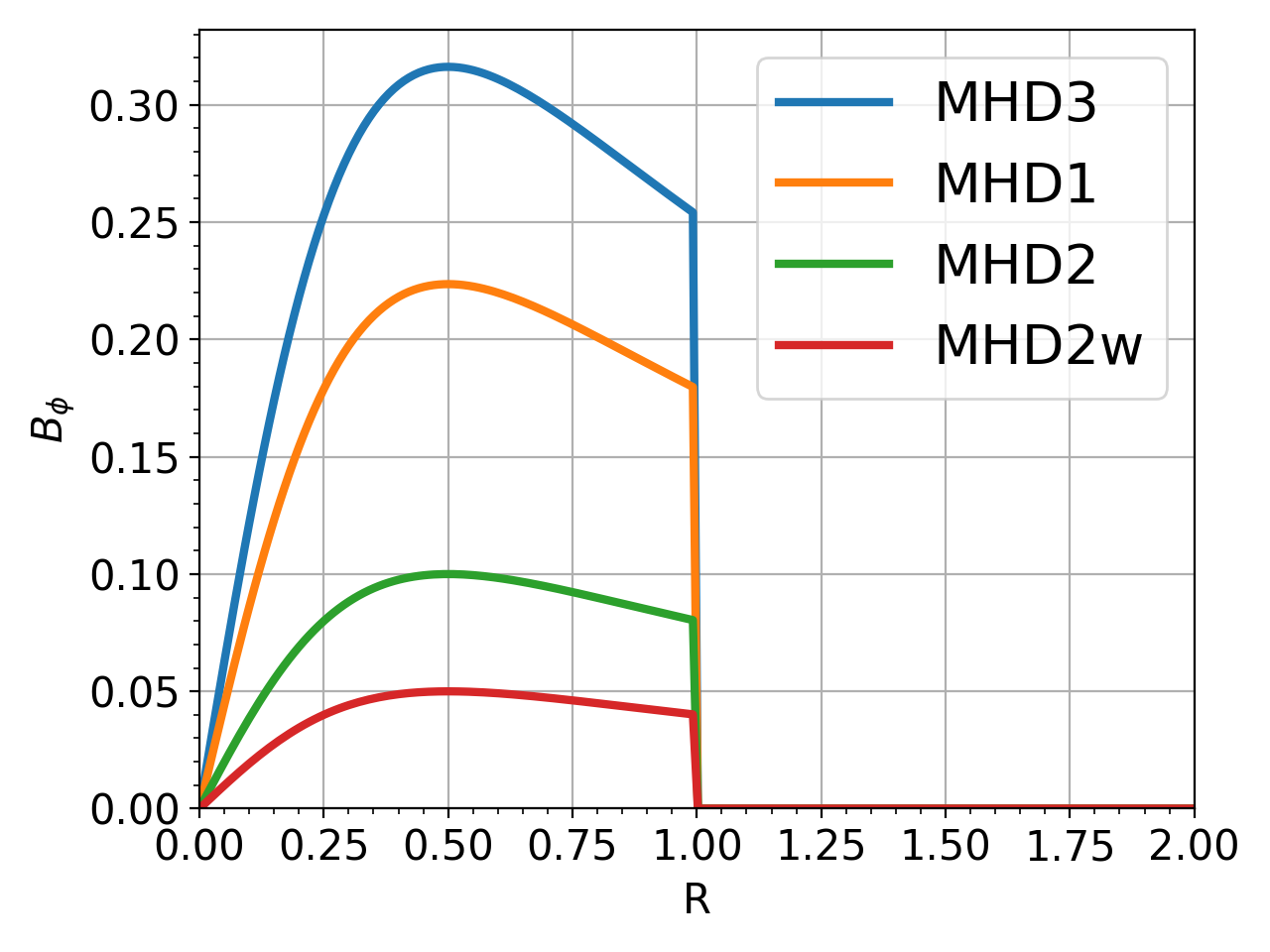}
    \includegraphics[width=0.8\columnwidth]{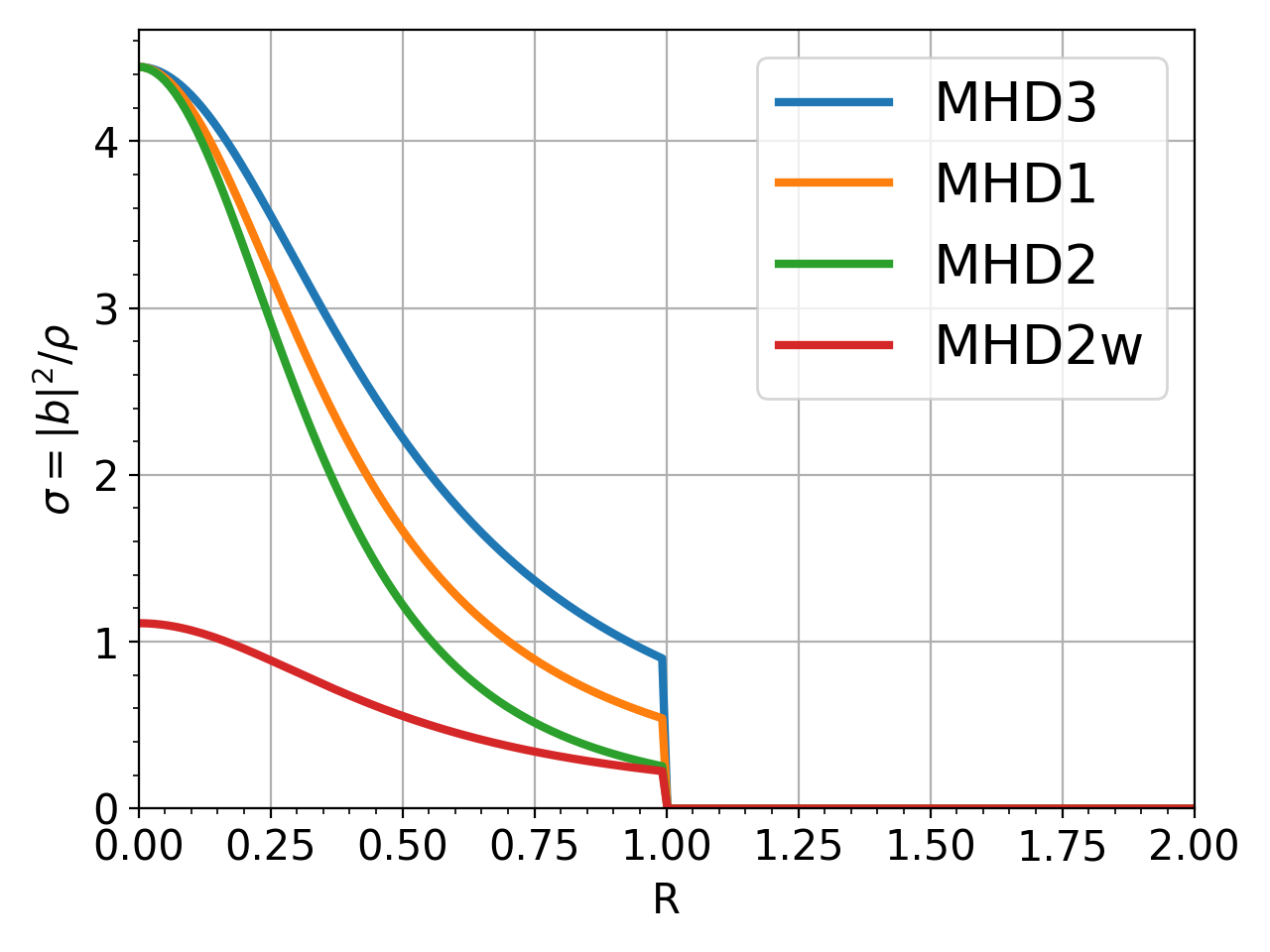}
    \caption{Initial distributions of toroidal fields(the left panel) and magnetizations(the right panel) for various cases.}
    \label{fig:init}
\end{figure}

Table~\ref{T1} shows the basic numerical setup and numerical methods that we used in our simulations.
As following previous work by \citet{mizuno2015}, we consider the cylindrical over-pressured jet to be filled in the simulation domain.
The jet radius is $R_j=1$. It has an axial velocity with Lorenz factor $\gamma_j=3$ which corresponds to $v_j=0.9428c$. 
The jet is lighter ($\rho_j=0.01$) and over-pressured ($p_j=1.5p_a$) than the ambient medium initially.
The ambient medium is stationary  ($v_a=0$) with constant density ($\rho_a=1$) and constant gas pressure ($p_a=0.018$). A preexisting cylindrical jet has a helical magnetic field. The poloidal and toroidal components of the magnetic field are given by
\begin{equation}
    B_z=\frac{B_0}{1+(R/a)^2},\qquad B_\phi=k\frac{B_0(R/a)}{1+(R/a)^2}, \label{5}
\end{equation}
where we use $B_0=0.2$ in fiducial models. $a$ and $k$ are parameters related to determining the magnetic pitch profile:
\begin{equation}
    P=\frac{R}{R_j}\left(\frac{B_z}{B_\phi}\right)=\frac{a}{k R_j}.\label{6}
\end{equation}
In our fiducial model ({\tt MHD1}), we use $a=0.5$ and $k=\sqrt{5}$ which leads to a magnetic pitch of $P=0.22$ and an average magnetisation $\langle \sigma \rangle=1.33$.
To explore the effect of magnetic pitch, we also simulate the cases with different magnetic pitches such as $k=1$ ({\tt MHD2}) and $\sqrt{10}$ ({\tt MHD3}) which correspond to a higher pitch ($P=0.5$), and a lower pitch ($P=0.16$) than our fiducial model, respectively. We should note that in our definition of magnetic pitch, higher (lower) pitch case has a more poloidal (toroidal) component of magnetic field dominates.
The case {\tt MHD2w} uses a weaker magnetic field ($B_0=0.1$) than the model {\tt MHD1} to discuss the effect of the magnetic field. Table~\ref{T2} lists the setups of various cases. 
The effective inertia is nearly invariant among different cases (see Sec.~\ref{dis} for details). 
Figure~\ref{fig:init} shows the initial radial distribution of toroidal magnetic field and magnetization in different cases. We set the jet radius as $R_j=1$. All quantities are sharply changed at the jet boundary. As you see from the figure, in most simulation models the jets are highly magnetized ($\sigma>1$).

\begin{table}[!h]
    \centering
    \begin{tabular}{lllll}
    \hline
    case         & $B_0$ & k           & P    & $\langle \sigma \rangle$ \\ \hline
    {\tt MHD1-2D} & 0.2   & $\sqrt{5}$  & 0.22 & 1.33\\
    {\tt HD}     & -     & -           & -    &  -\\
    {\tt MHD1}   & 0.2   & $\sqrt{5}$  & 0.22 & 1.33\\
    {\tt MHD2}   & 0.2   & $1$         & 0.50 & 0.98 \\
    {\tt MHD3}  & 0.2   & $\sqrt{10}$ & 0.16 & 1.79 \\
    {\tt MHD2w}   & 0.1   & $1$         & 0.50 & 0.24\\
    \hline
    \end{tabular}
    \caption{Basic properties of the different cases, where  $\langle \sigma \rangle$ is averaged magnetization.}
    \label{T2}
\end{table}

The computational domain is $(x,y,z)=(\pm 5R_j,\pm5R_j,60R_j)$ with a uniform grid of $(N_x,N_y,N_z)=(500,500,600)$. This corresponds to 50 cells per jet radii in transversal and 10 cells per jet radii in vertical direction.
We impose outflow boundary conditions on all surfaces except for $z=z_{min}$. At $z=0$, we use fixed boundary conditions that continuously inject the over-pressured jet into the computational domain. All simulations are performed up to $t_s=400$, where the simulation time unit is $t_s=R_j/c$.

\begin{figure*}[!ht]
    \centering
        \subfigure[]{\includegraphics[width=0.18\textwidth]{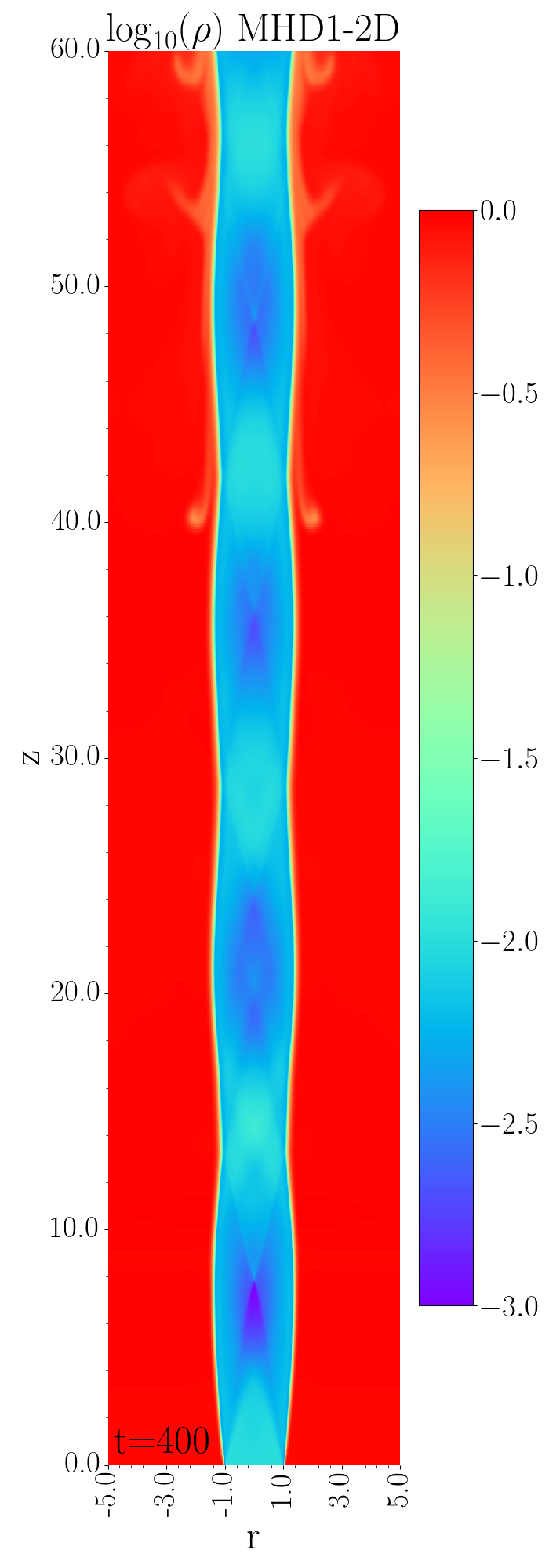}}
        \subfigure[]{\includegraphics[width=0.18\textwidth]{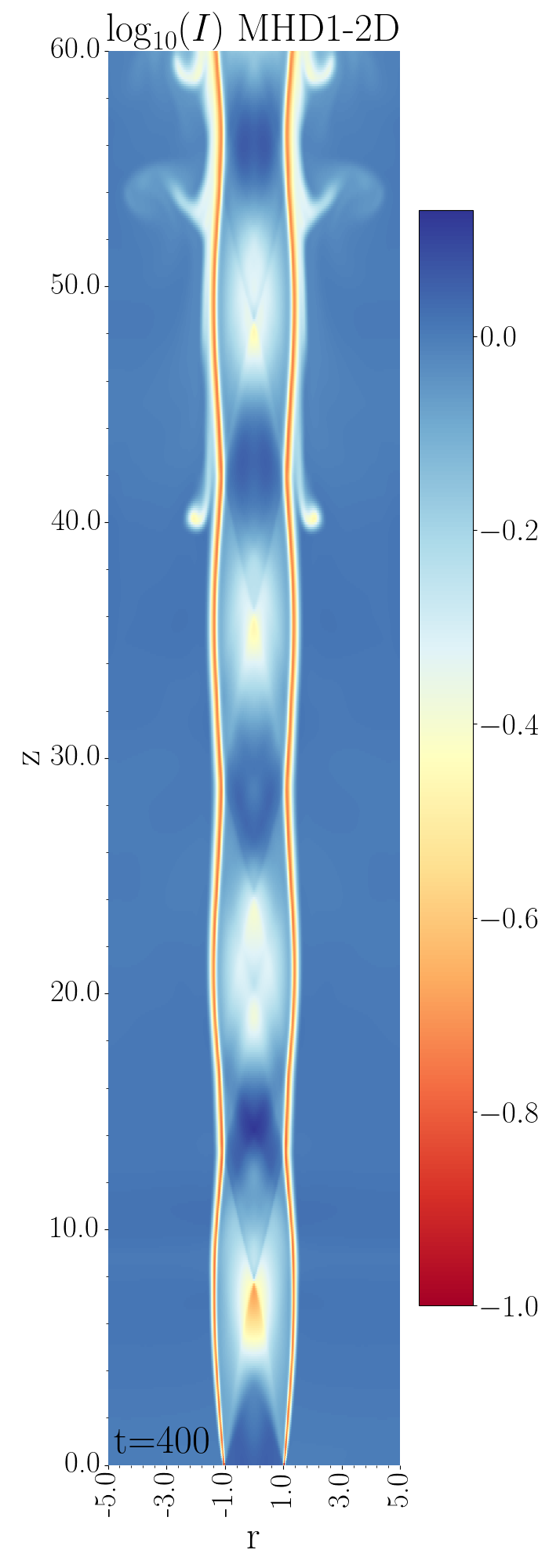}}
        \subfigure[]{\includegraphics[width=0.18\textwidth]{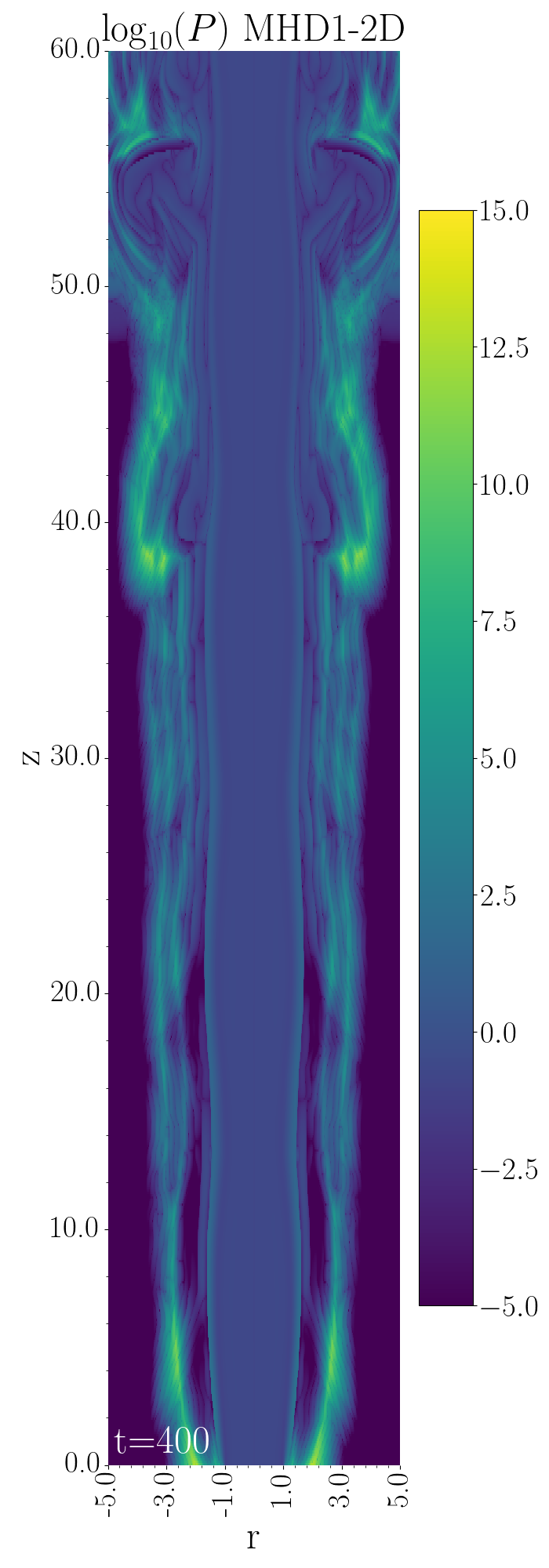}}
    \caption{Distribution of logarithmic density (left), effective inertia, $I=\gamma^2\rho h+B_z^2+B_\phi^2$ (middle), and magnetic pitch (right) for 2D RMHD model {\tt MHD1-2D} at $t_s=400$.
    }
    \label{fig:2D}
\end{figure*}

\section{Results}\label{re}
\subsection{2D helically-magnetized jet}\label{2d}

First, we present the 2D RMHD simulations of over-pressured jets with a helical magnetic field.
For the 2D simulation ({\tt MHD1-2D}), we use a cylindrical coordinate $(R,z)$, and the axisymmetric boundary is employed at $R=0$.
The simulation domain is $[R,z]=[5R_j, 60R_j]$ with grid number of $(N_R,N_z)=(250, 600)$.

Figure~\ref{fig:2D} shows 2D distributions of the rest-mass density, effective inertia ($I=\gamma^2\rho h+B_z^2+B_\phi^2$), and magnetic pitch for the 2D RMHD case at $t_s = 400$. As pointed out by \cite{Matsumoto2013}, effective inertia is a key quantity to understand the stability of recollimation shock. Here, we present the effective inertia in the radial direction. The effective inertia is useful to investigate the recollimation shock structure analytically. A detailed discussion is provided in Appendix~\ref{acce}.

Due to the jet being over-pressured than the external medium initially, from the simulation begins, shocks and rarefaction waves are created at the boundary of the jet and the external medium. These propagate inside and outside the jet to make a recollimation shock structure as seen in \cite{mizuno2015}.
The jet appears to be a stable recollimation shock structure as shown in Figure~\ref{fig:2D}. The separation of recollimation shocks is mostly constant due to the constant external medium.
The plot of the effective inertia (the middle panel) shows a low inertia interface between the jet and the ambient environment. This is because the gas at the interface is less dense than the environment and has a lower Lorentz factor than the inner jet. Meanwhile, the right panel of Figure~\ref{fig:2D} illustrates that the magnetic pitch also forms a boundary at the interface and the outer part shows a hierarchical structure. Some peculiar structures are visible outside the jet at z>40, which are components ejected by the jet during the establishment of recollimation structures. Thus, these are kind of numerical artifacts. Similar to \cite{mizuno2015}, in 2D simulations, we do not see any excitement of the instabilities in jets. From the analytical approach, we can mostly reproduce the radial motion of the recollimation shock structure (see Appendix~~\ref{acce}).

\begin{figure*}[!ht]
    \centering
    \subfigure[]{\includegraphics[width=0.18\textwidth]{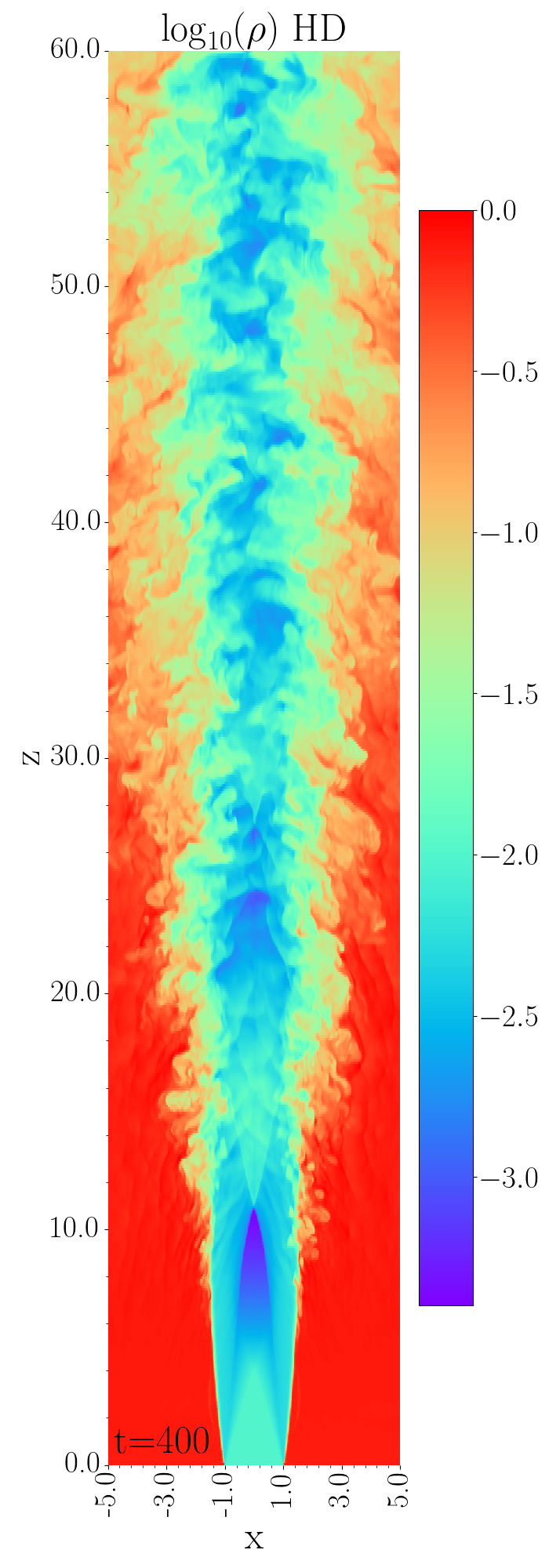}}
    \subfigure[]{\includegraphics[width=0.18\textwidth]{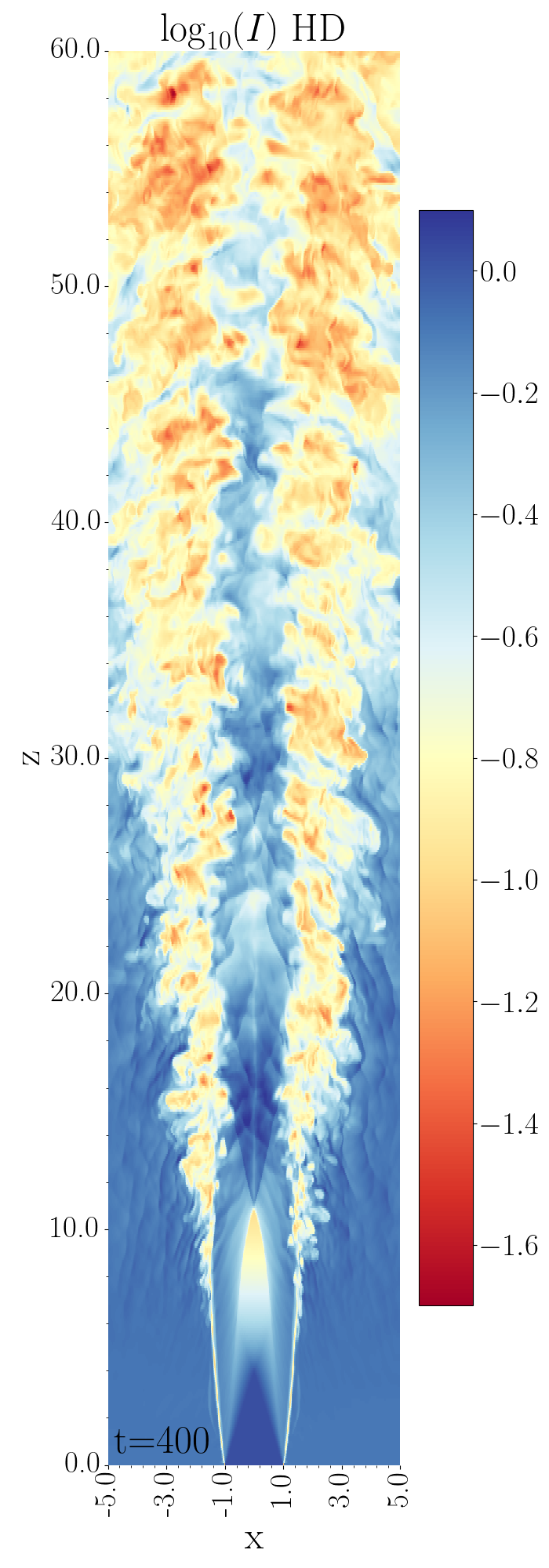}}
    \subfigure[]{\includegraphics[width=0.18\textwidth]{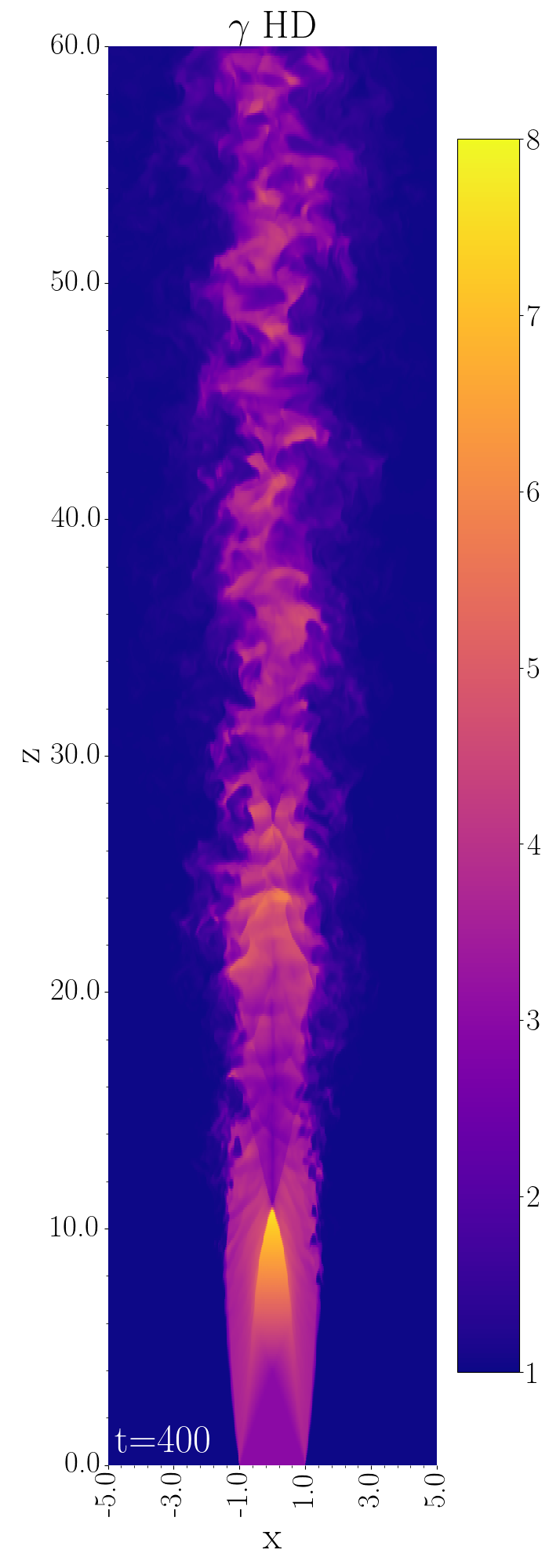}}
    \caption{2D axial distribution of logarithmic density ({\it a}), effective inertia, $I=\gamma^2\rho h$ ({\it b}), and the Lorentz factor ({\it c})  for the 3D RHD case {\tt HD} at $t_s=400$.}
    \label{fig:3D HD}
\end{figure*}

\subsection{3D hydrodynamic jet}\label{3dhd}

\begin{figure}[!h]
    \centering
    \includegraphics[width=0.9\columnwidth]{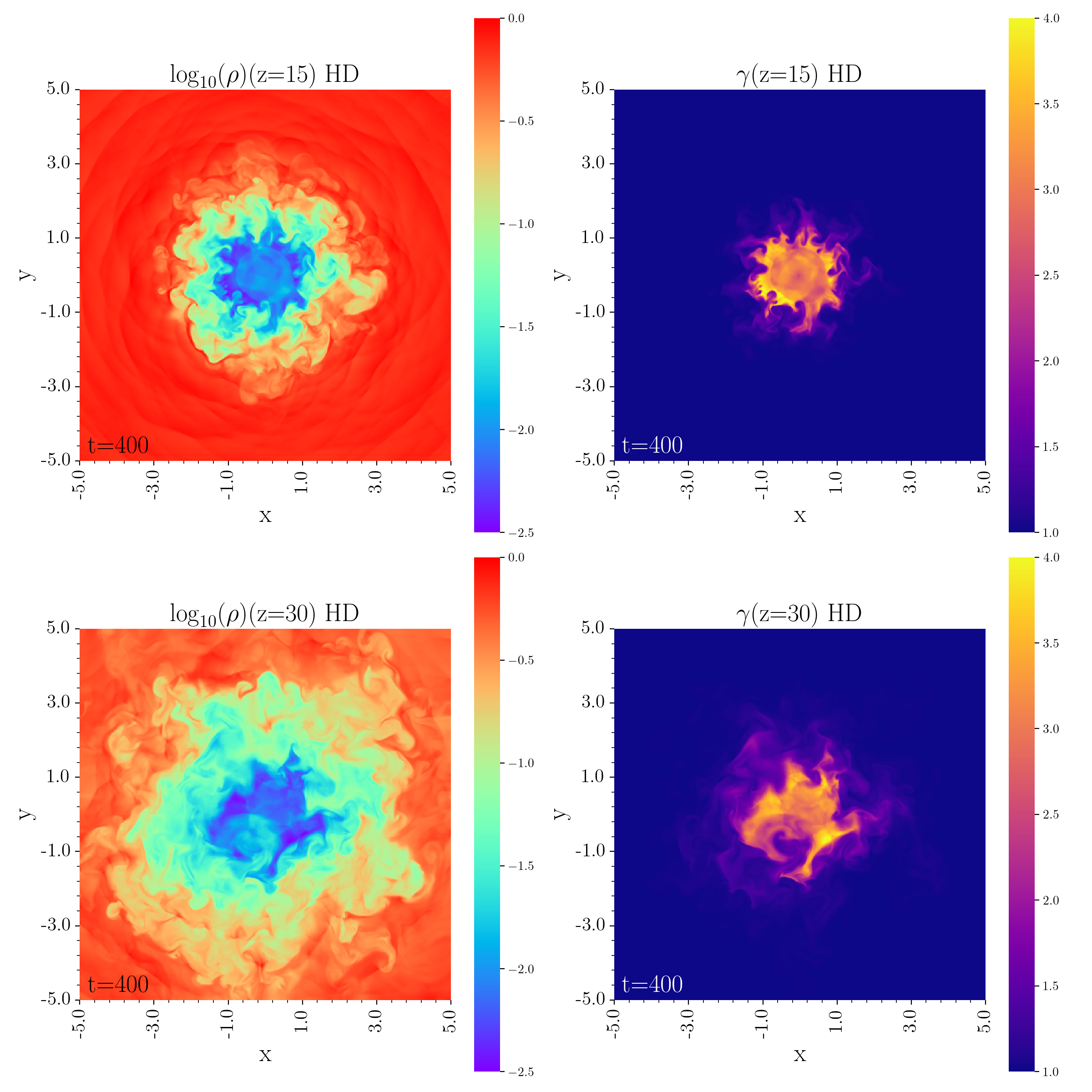}
    \caption{Distribution on xy plane of density ({\it left}) and Lorentz factor ({\it right}) at $z=15\,R_j$ ({\it top}) and $z=30\,R_j$ ({\it bottom}) for {\tt HD} case at $t_s=400$.}
    \label{fig:HDcut}
\end{figure}

We performed 3D RHD simulations of an over-pressure jet that is marked as the case {\tt HD}. The results are shown in figure~\ref{fig:3D HD}. From the density distribution, we clearly see the development of instability at the jet boundary. After creating the first recollimation shock structure at $z=10\,R_j$, the jet begins to transform into turbulence by growing the instability before forming a complete recollimation structure. Figure~\ref{fig:HDcut} shows the distribution on the xy plane of density and Lorentz factor at two different axial positions that indicate the development of RTI at different distances from the inlet. At $z=15\,R_j$ (top panels), we see finger-like structures at the boundary between the jet and ambient medium, which are typical characteristics of RTI. In the further distance from the jet inlet ($z=30\,R_j$), RTI is well developed and tends to non-linear evolution which leads to the turbulent structure with the mixing layer of jet and ambient medium. From the Lorentz factor plots, we see the jet is accelerated during this development.

In our 3D RHD simulation, the initial effective inertia of jet $I_j=\gamma^2\rho h=1.062$ is slightly smaller than the ambient environment ($I_a=1.072$). Our 3D RHD simulation satisfies the stability condition by \cite{Matsumoto2013}. However, later \cite{Matsumoto2017} have provided an updated analytical stability criterion for RTI in the following form:
\begin{equation}
    \left[\gamma^2\rho(1+\frac{\Gamma^2}{\Gamma-1}\frac{p_{gas}}{\rho})\right]_{\rm j} < \left[\gamma^2\rho(1+\frac{\Gamma^2}{\Gamma-1}\frac{p_{gas}}{\rho})\right]_{\rm a}, \label{7}
\end{equation}
which is slightly different from the definition of effective inertia, $I=\gamma^2 \rho h$.
Based on this updated analytical stability criteria, our simulation condition does not satisfy this stability condition, which indicates RTI can grow in our 3D RHD simulation. The prediction of the criterion is consistent with the simulation. 



\subsection{3D helically magnetized jet}\label{3d}
\begin{figure*}[!ht]
    \centering
    \subfigure[]{\includegraphics[width=0.18\textwidth]{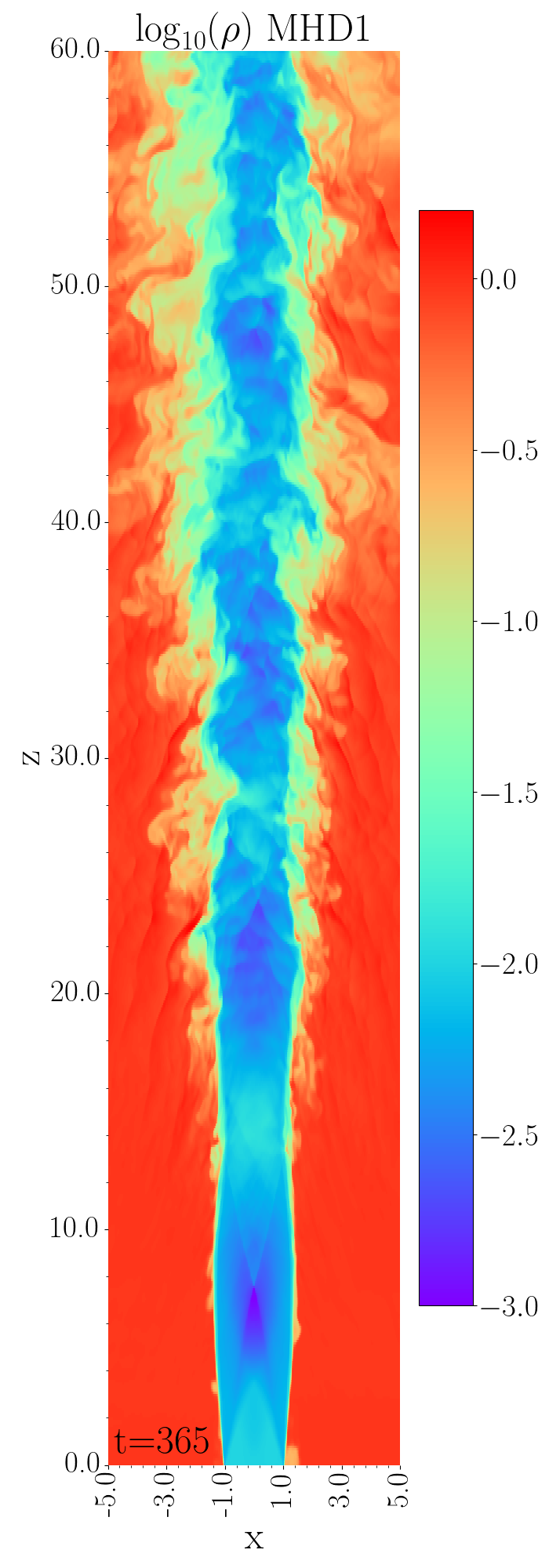}}
    \subfigure[]{\includegraphics[width=0.18\textwidth]{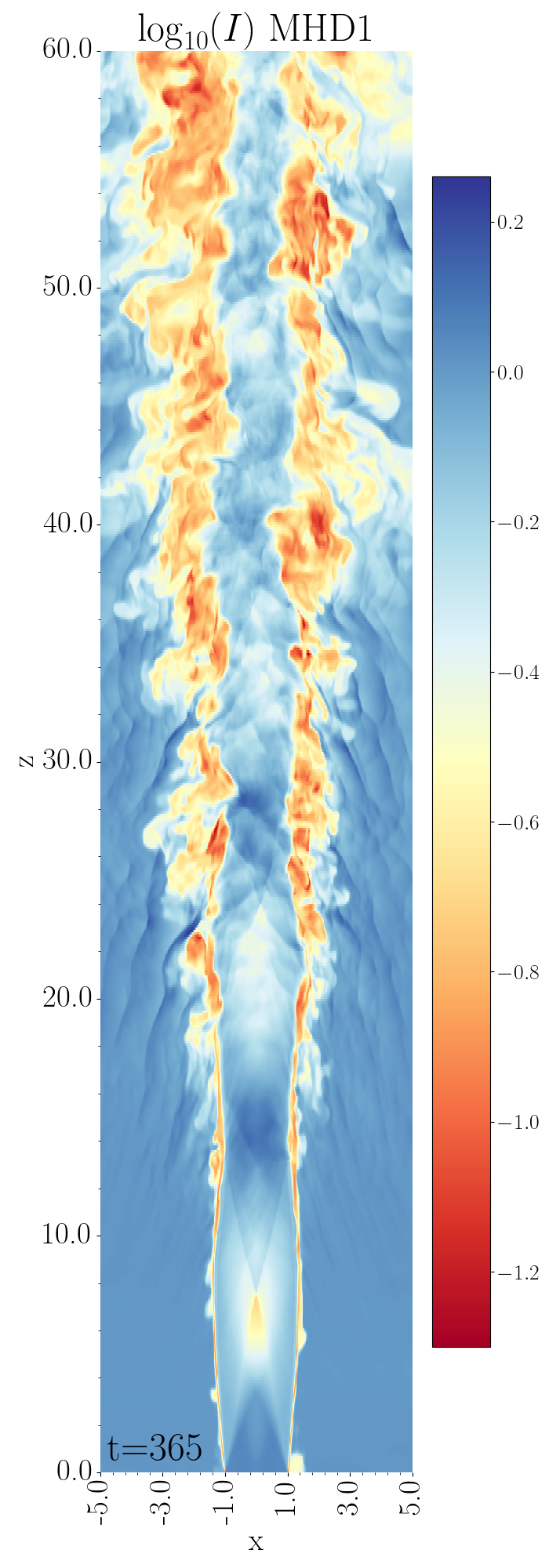}}
    \subfigure[]{\includegraphics[width=0.18\textwidth]{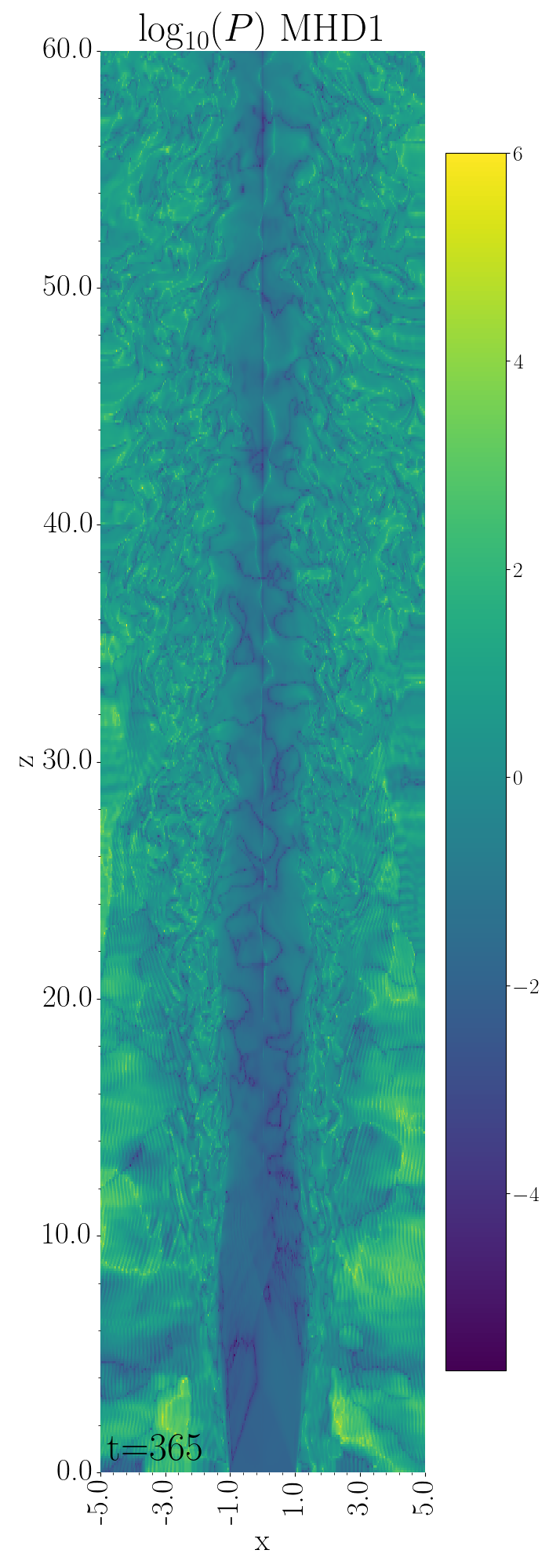}}
    \subfigure[]{\includegraphics[width=0.18\textwidth]{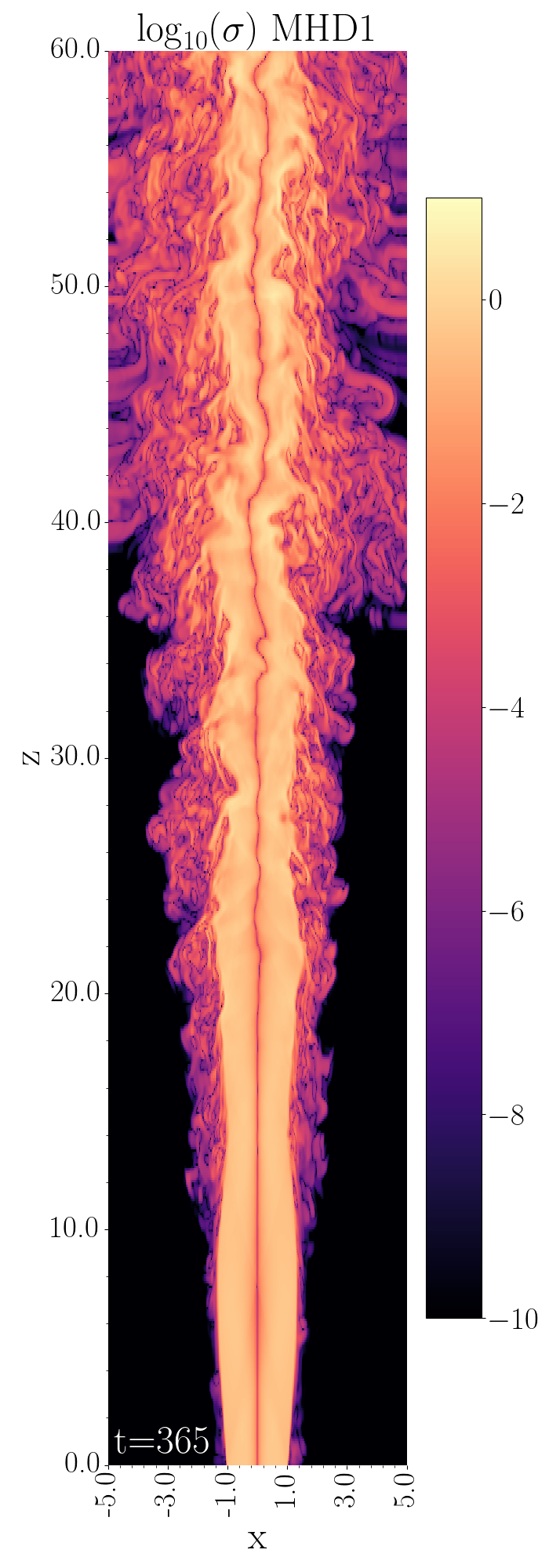}}
    \subfigure[]{\includegraphics[width=0.18\textwidth]{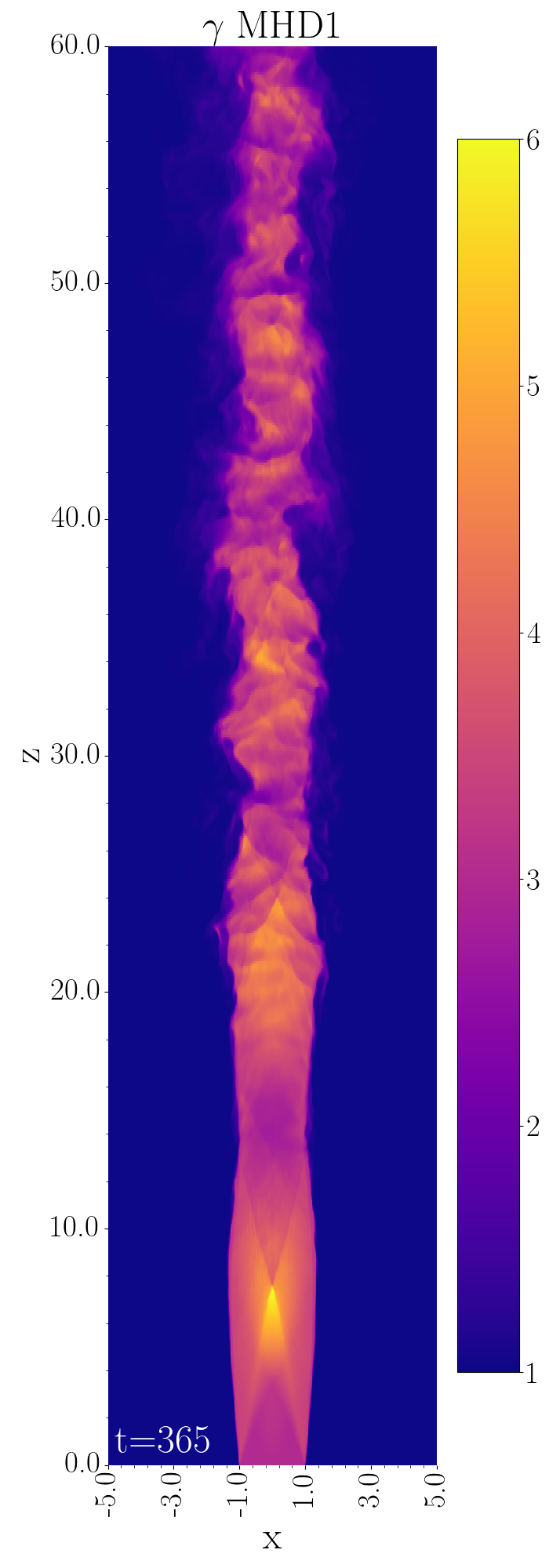}}
    \caption{2D axial distribution of logarithmic density ({\it a}), effective inertia ({\it b}), magnetic pitch ({\it c}), the magnetization ({\it d}), and Lorentz factor ({\it e}) for the 3D RMHD case {\tt MHD1} at $t_s=365$.}
    \label{fig:3D}
\end{figure*}

In the previous subsection, we have shown that recollimation shock structure is stable in 2D axisymmetric RMHD simulations of over-pressured jets with a helical magnetic field. However, in 3D RHD simulations, we see the development of RTI. Next, we perform 3D RMHD simulations of over-pressured jets with a helical magnetic field.

\begin{figure}[!h]
    \centering
    \includegraphics[width=0.8\columnwidth]{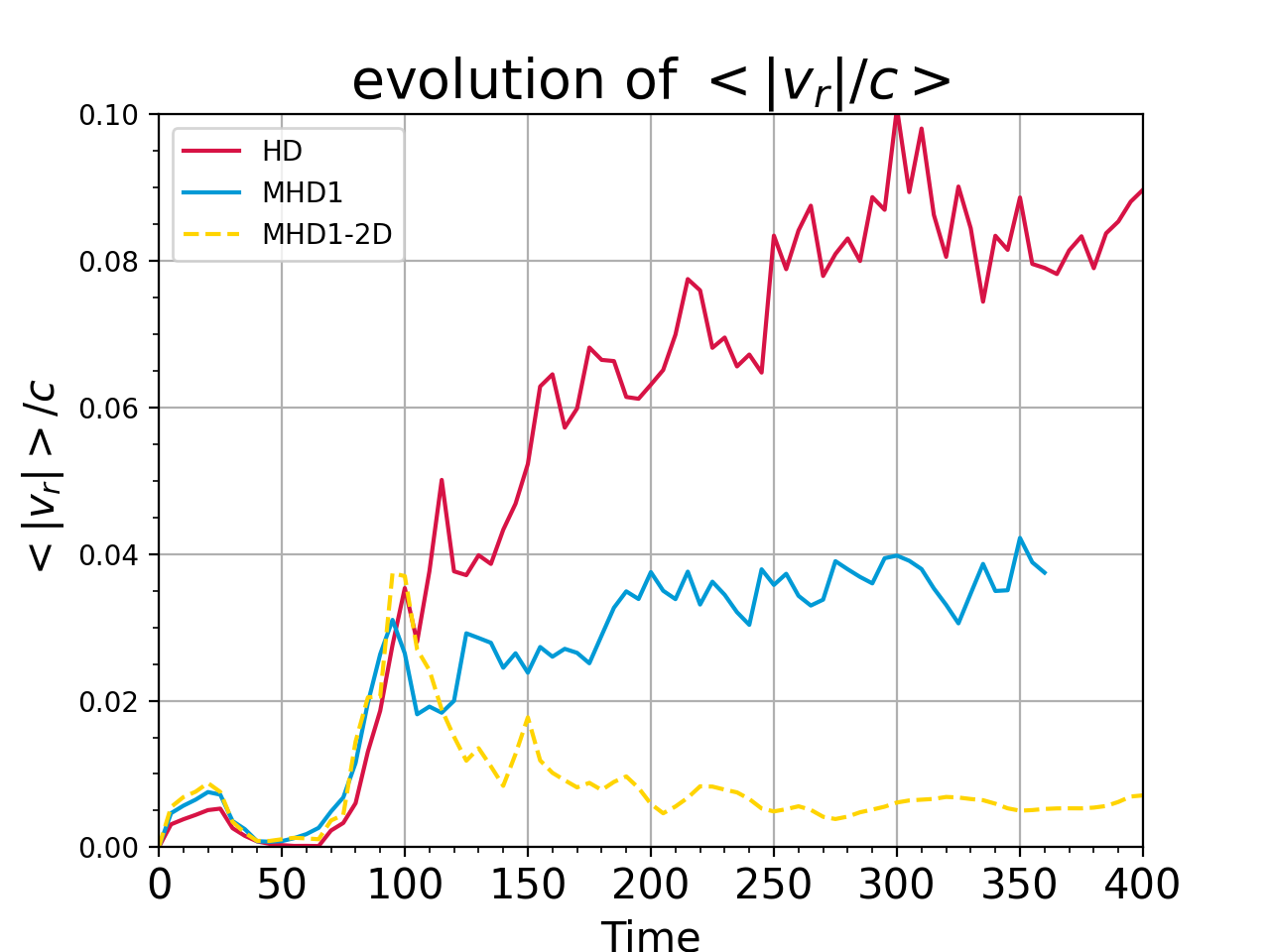}
    \caption{The time evolution of averaged radial velocity $|v_r|/c$ at $z=60\,R_j$ with the cases of 3D RHD ({\tt HD}, red), 2D RMHD ({\tt MHD1-2D}, yellow-dashed), and 3D RMHD ({\tt MHD1}, blue).
    }
    \label{fig:vr}
\end{figure}

The fiducial model takes the same numerical setup as the 2D RMHD simulation shown in Section~\ref{2d}. 
In Figure~\ref{fig:3D}, we show the distributions of density, effective inertia, and magnetization $\sigma$ of the {\tt MHD1} at $t_s=365$, where the form of magnetization $\sigma=B_\phi^2/\gamma^2\rho$ that was used in \cite{millas2017}. 
We should note that unfortunately, the simulation was stopped at $t=365$ due to numerical reasons. However, the development of RTI is saturated around $t_s=250$ (see figure~\ref{fig:vr}). Thus we think the distribution at $t_s=365$ is still comparable with other cases shown at $t_s=400$.

In the comparison with the 2D RMHD case, we clearly see the development of the RTI that is triggered in the 3D simulation. In the region far from the jet inlet, RTI is fully developed, and the turbulent structure is excited.
Moreover, from the plot of effective inertia, we can observe that the low inertia interface still exists in the turbulent region and it also develops finger-like structures, separating inner and outer turbulence. Unlike the 2D case, the magnetic pitch in the 3D case shows a complex structure.
As \cite{Matsumoto2017} indicated, the difference in effective inertia between the jet and the ambient medium does not give a criterion for the growth of the RTI. However, it is still useful to trace the recollimation shock structure (see Appendix~\ref{acce}).

The panel (c) of Figure~\ref{fig:3D} implies the RTI is triggered even when the region with high magnetization $\sigma >10$. It seems a contradiction to the statement that higher magnetization ($\sigma>0.01$) stabilizes the jet suggested by \cite{millas2017}. However, in their simulations, they used a pressure continuity condition at the boundary while we considered over-pressure jets. This may indicate that it would be difficult to derive a quantitative stability criterion due to the complex properties of the jet. 

\begin{figure}[!h]
    \centering
    \includegraphics[width=0.9\columnwidth]{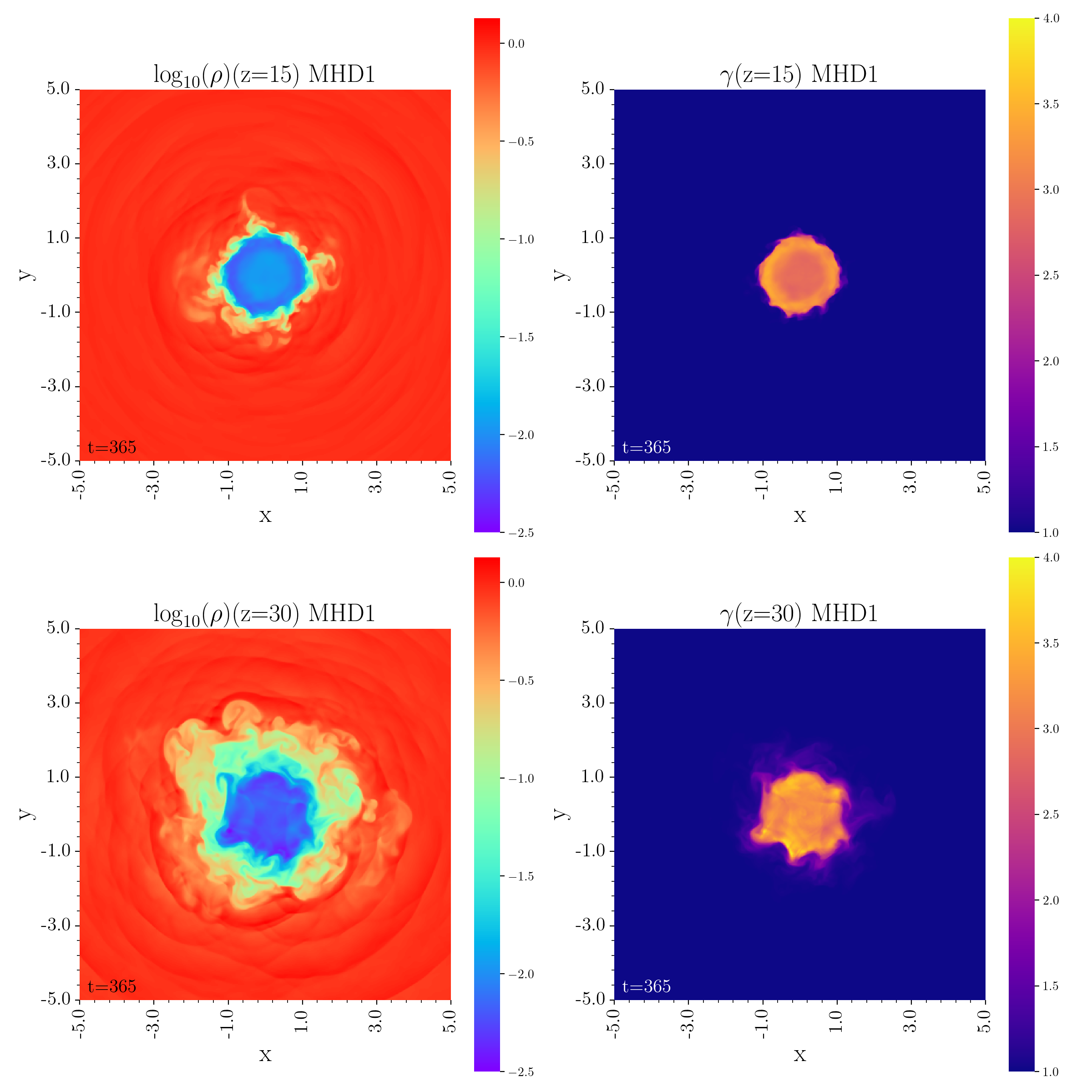}
    \caption{Distribution on xy plane of density and Lorentz factor at $z=15\,R_j$ (top) and $z=30\,R_j$ (bottom) for {\tt MHD1} case at $t_s=365$.}
    \label{fig:MHDcut}
\end{figure}

Similar to Figure~\ref{fig:HDcut}, Figure~\ref{fig:MHDcut} shows the distribution on the xy plane of density and Lorentz factor at two different axial positions that indicate the development of RTI at different distances from the jet inlet. Obviously, the existence of the toroidal magnetic field slows down the development of RTI.


To quantitatively analyze the strength of RTI among different cases, we plot the time evolution of average radial velocity $|v_r|/c$ at $z=60\,R_j$ for the {\tt HD}, {\tt MHD-2D}, and {\tt MHD1} cases, respectively in Fig.~\ref{fig:vr}. The recollimation shock starts to develop on the top of the simulation box at $t_s=64$ and a jet reaches a fully developed state at $t_s=100$. After that, stable recollimation shock structures are formed in the 2D RMHD case ({\tt MHD-2D}) that indicates the $\langle |v_r|/c \rangle \sim 0$. However, for the 3D RHD case ({\tt HD}), the RTI starts to grow after recollimation shock develops. Thus, the averaged radial velocity grows continuously in later simulation time due to the development of turbulent structure. For the 3D RMHD case ({\tt MHD1}), the magnetic field, in particular the toroidal field eliminates the development of RTI. It is indicated that the strength of average radial velocity becomes smaller than that in {\tt HD} case. We note that axial variation by the development of instability depends a little on the spatial resolutions. However, the results do not change it.
%

\subsection{Effects of magnetic strength and pitches}\label{effect}
\begin{figure*}[!ht]
    \centering    
    \subfigure[]{\includegraphics[width=0.18\textwidth]{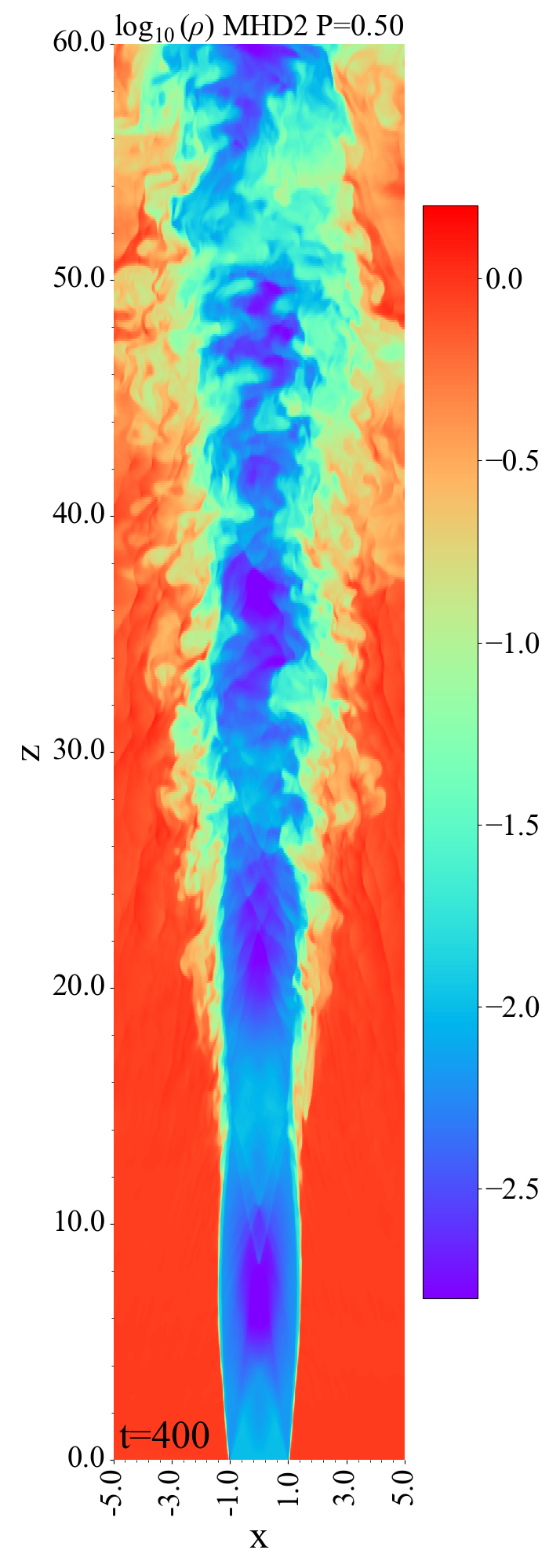}}
    \subfigure[]{\includegraphics[width=0.18\textwidth]{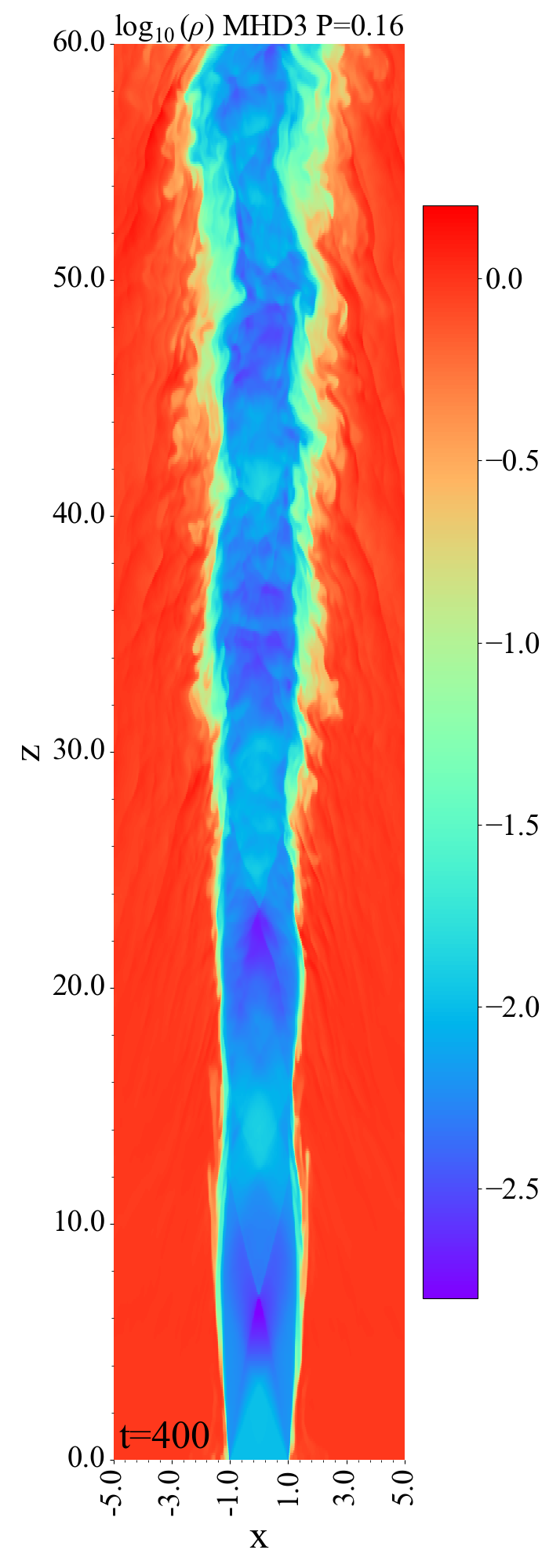}}
    \subfigure[]{\includegraphics[width=0.18\textwidth]{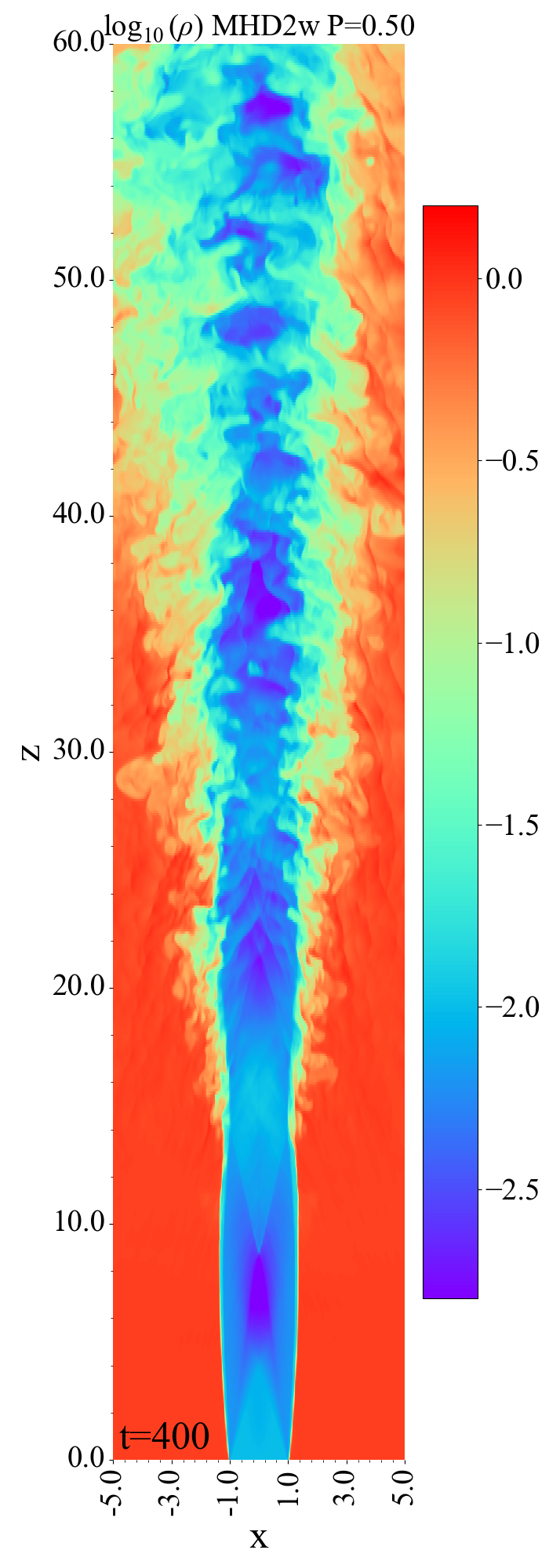}}
    \caption{2D Axial distribution of logarithmic density for the cases of different magnetic pitches and magnetic strength at $t_s=400$. ({\it a}) higher magnetic pitch case ({\tt MHD2}) with $k=1$, $P=0.50$, ({\it} b) lower magnetic pitch case ({\tt MHD3}) with $k=\sqrt{10}$, $P=0.16$, and ({\it c}) lower magnetization case ({\tt MHD2w}) with $B_0=0.1$, $k=1$, and $P=0.50$.}
    \label{fig:k}
\end{figure*}
\begin{figure}[!h]
    \centering
    \includegraphics[width=0.9\columnwidth]{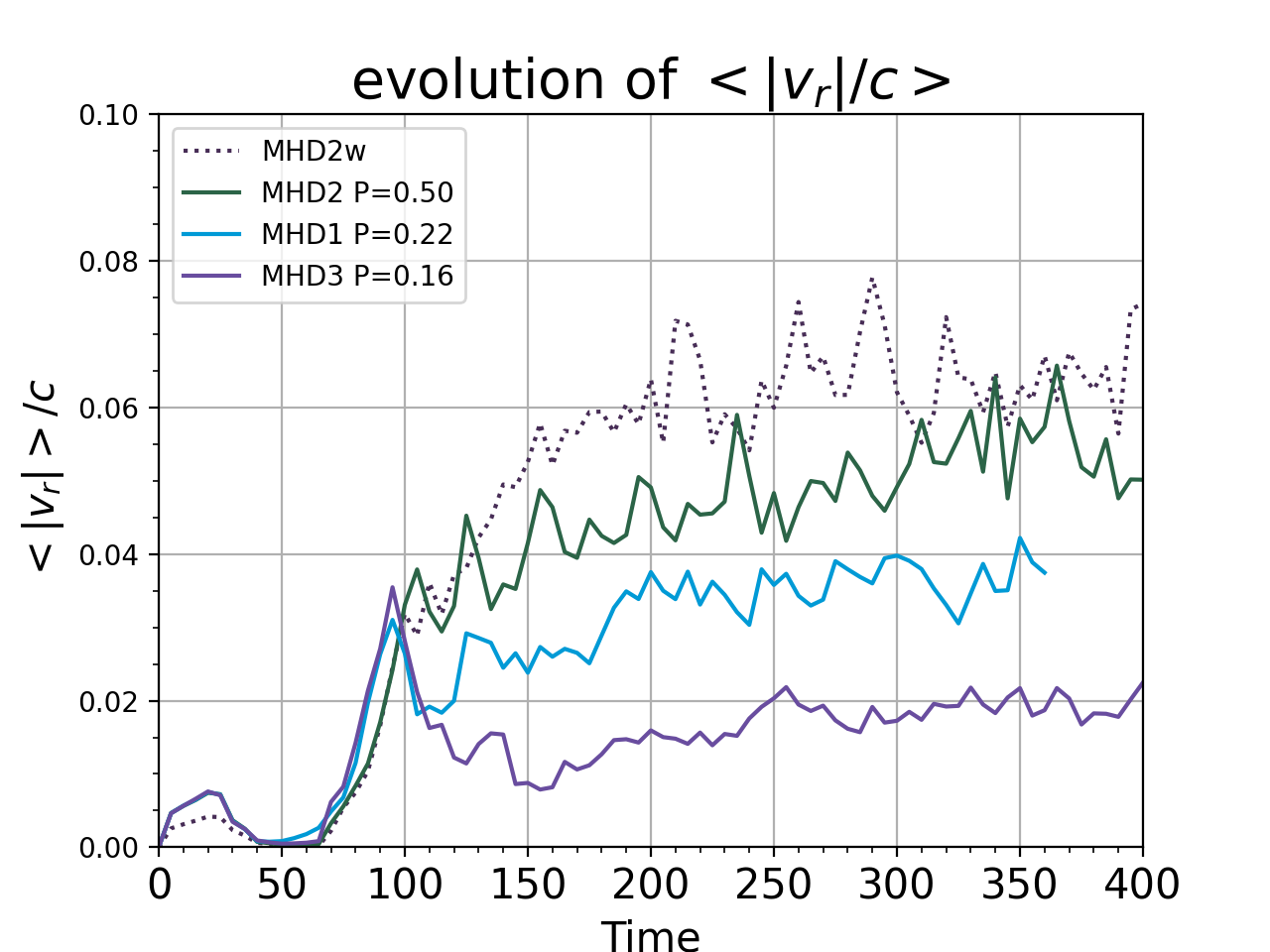}
    \caption{Same as figure~\ref{fig:vr} but for cases with different magnetic pitches. The green line represents the higher magnetic pith case {\tt MHD2} ($k=1$, $P=0.5$) and the purple line indicates the lower magnetic pitch case {\tt MHD3} case ($k=\sqrt{10}$, $P=0.16$). For a reference, we plot the representative {\tt MHD1} case ($k=0.5$, $P=0.22$) as blue line. Dotted line represents the weaker magnetized case {\tt MHD2w} ($B_0=0.1$, $k=1$, $P=0.5$).
    }
    \label{fig:vr_k}
\end{figure}

We continue our investigation by comparing the simulation results with different magnetic pitch cases, lower magnetic pitch ({\tt MHD2}) with $P=0.16$ and higher magnetic pitch ({\tt MHD3}) with $P=0.5$. 
Figure~\ref{fig:k} shows the density distributions at $t_s=400$.

When magnetic pitch becomes higher ({\tt MHD2}; left panel of Fig.~\ref{fig:k}), the RTI grows faster. Thus, the jet is more disrupted in the further region from the jet inlet than in the case of {\tt MHD1}.
On the other hand, with the magnetic pitches decreasing (case {\tt MHD3}; middle panel of Fig.~\ref{fig:k}), the RTI is mostly prohibited as is also indicated by \citet{millas2017,matsumoto2021}. 

Fig.~\ref{fig:vr_k} shows the time evolution of averaged radial velocity at $z=60\,R_j$ for three different magnetic pitch cases. From this figure, we clearly see that the lower pitch case (green line) has a larger averaged radial velocity than our fiducial model (blue line). When the magnetic pitch becomes higher (purple line), the averaged radial velocity significantly reduces. Thus, magnetic pitch is crucial for the growth of RTI.

As seen in Fig.~\ref{fig:vr}, the existence of a magnetic field reduces the development of RTI. When the magnetic field becomes weaker, the growth of RTI becomes stronger as is seen in the right panel of Fig.~\ref{fig:k}. From the time evolution of averaged radial velocity (dotted line of Fig.~\ref{fig:vr_k}), the weaker RMHD case {\tt MHD2w} is properly between the {\tt MHD1} case and the {\tt HD} case (see in Fig.~\ref{fig:vr}). It indicates that magnetic field strength is an important factor for the development of RTI.

When the RTI no longer fully disrupts the jet structure in higher magnetic pitch cases such as the {\tt MHD3} case (the middle panel in Fig.~\ref{fig:k}), we can see the helically twisted structure in the jets.
It implies the evidence of growing current-driven kink instability \citep[e.g.,][]{mizuno2014,singh2016}. We are going to discuss it in the following Section~\ref{kink}.

\subsection{possibility of other instabilities}\label{other}

\begin{figure}[!h]
    \centering
    \includegraphics[width=0.45\columnwidth]{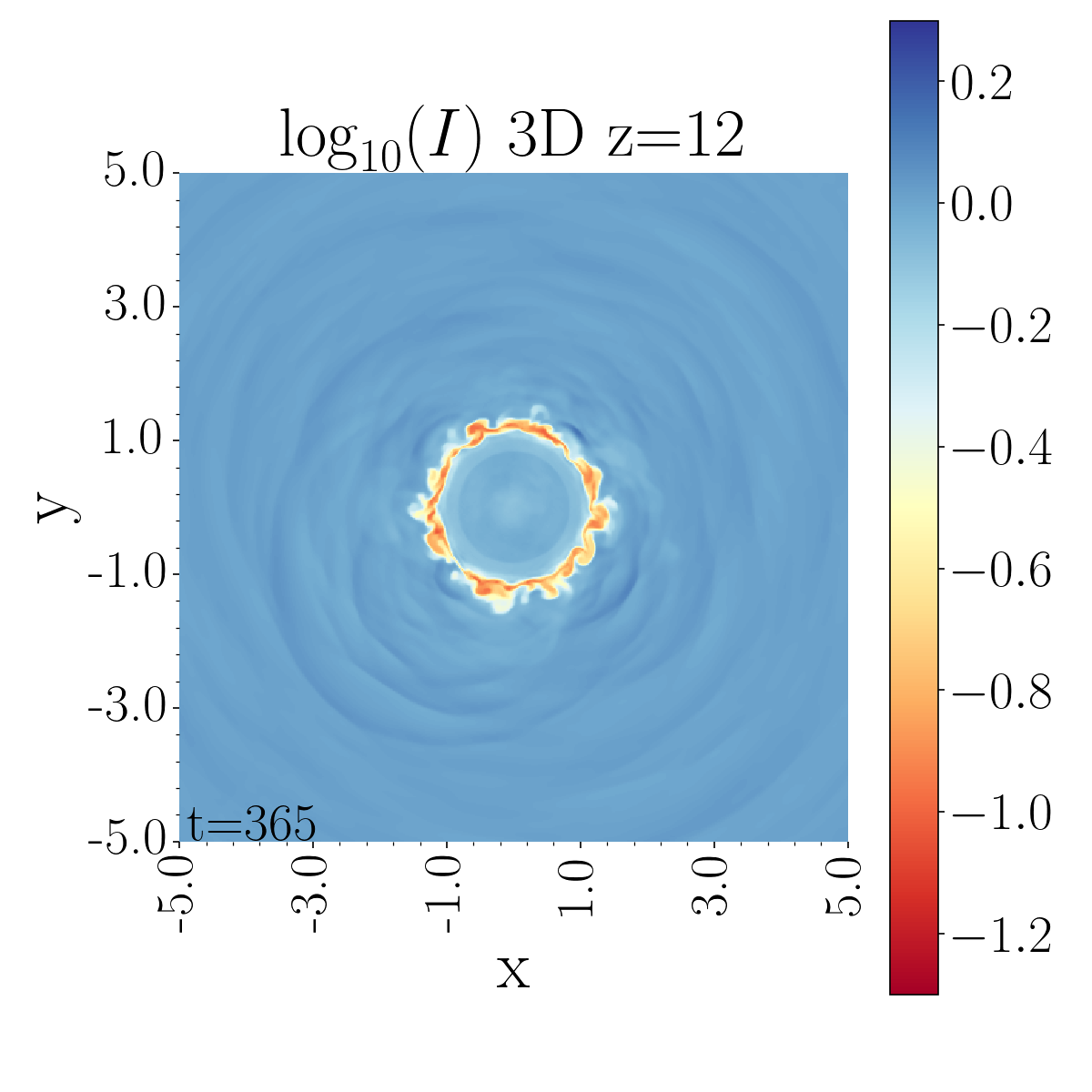}
    \includegraphics[width=0.45\columnwidth]{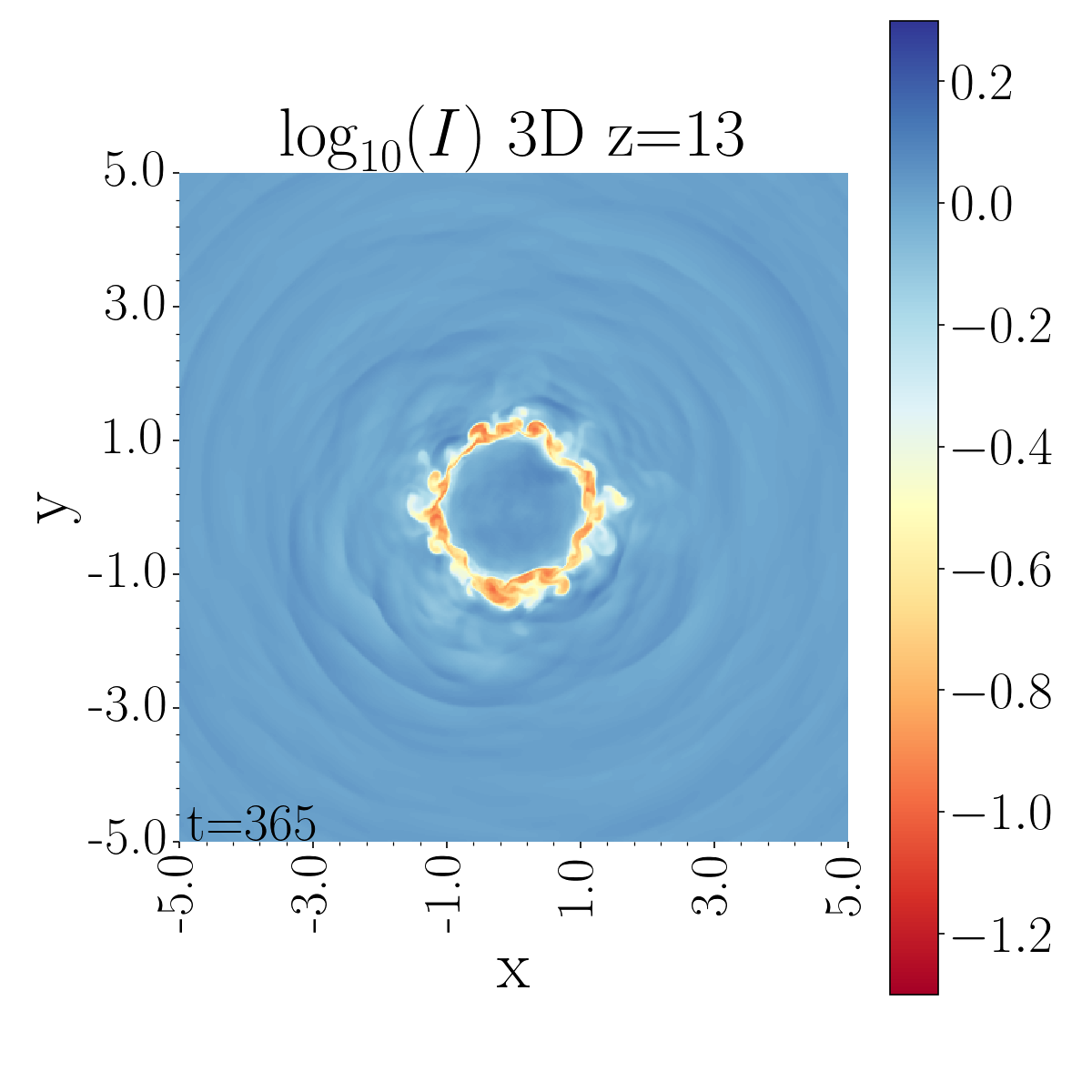}
    \caption{Distributions on xy plane of effective inertia at $z=12R_j$ (the left panel) and $z = 13 R_j$ (the right panel) for {\tt MHD1} case at $t_s = 365$.}
    \label{fig:RMI}
\end{figure}

In our 3D MHD simulation of over-pressured jet, we have seen the growth of instability at the boundary between jet and external medium that disrupts the jet structure. From the finger-like structure seen at the boundary and condition, we concluded it is due to the growth of RTI. However, other instabilities such as KHI, Richtmyer–Meshkov instability (RMI), and CFI can also contribute to the disruption of the jet except for RTI. In this sub-section, we evaluate the possibility of each instabilities.

For our simulations, KHI is mostly ruled out because initially, we do not have such a velocity shear boundary. We note that RTI is some kind of KHI.

RMI may be excited when the reflective shock from the jet axis reaches the contact continuity at the boundary. For {\tt MHD1} case, as is shown in the distribution of effective inertia at xy plane of Figure~\ref{fig:RMI}, reflective shock is observed around $z=12$. However there is no clear evidence of the growth of RMI what seen in \cite{Matsumoto2013} when the shock reaches the contact discontinuity.

From the previous studies by \cite{Komissarov2019} and \cite{matsumoto2021}, CFI is possible to growth in relativistic magnetized jets. However, it is also known that a weak magnetic field can inhibit it. Based on the definition of magnetization $\sigma_{\rm m}={b_\phi^2}/{\rho h}$, \cite{matsumoto2021} mentioned that when the jet magnetization becomes larger than 0.01, the CFI is stabilized. Under this definition, we calculate the maximum magnetizations for our MHD cases. These are $0.047$ ({\tt MHD1}), $0.0094$ ({\tt MHD2}), $0.094$ ({\tt MHD3}) , and $0.0023$ ({\tt MHD2w}), respectively. Most of our cases have a higher magnetization than the criteria suggested by \citet{matsumoto2021} . Thus, we do not need to consider that the developed turbulent structure of jet is caused by the growth of CFI.

Matsumoto et al. (2021) has proposed a more elaborate criterion, that is
\begin{equation}
    \frac{\sigma}{1+\sigma} > \frac{(\theta_0\gamma)^2}{16}, \label{eq:CFI}
\end{equation}
where $\theta_0$ is jet opening angle. To calculate the jet opening angle, we estimate the jet radii through the approach described in Appendix~\ref{acce} then we calculate the initial half-opening angle of jet. In the {\tt MHD1}, {\tt MHD2}, {\tt MHD3}, {\tt MHD2w} cases, the open angle is $0.093$, $0.074$, $0.082$, and $0.031$. The required magnetization for stabilizing the CFI is $\sigma> 0.0049$, $0.0031$, $0.0038$, and $0.0064$, respectively. Except {\tt MHD2w} case, the jet should be stable to CFI.

From these arguments, we conclude that the developed turbulence at the interface between jet and external medium is caused by the development of RTI.

\subsection{kink instability in the simulations}\label{kink}
\begin{figure*}[!ht]
    \centering
    \includegraphics[width=0.9\textwidth]{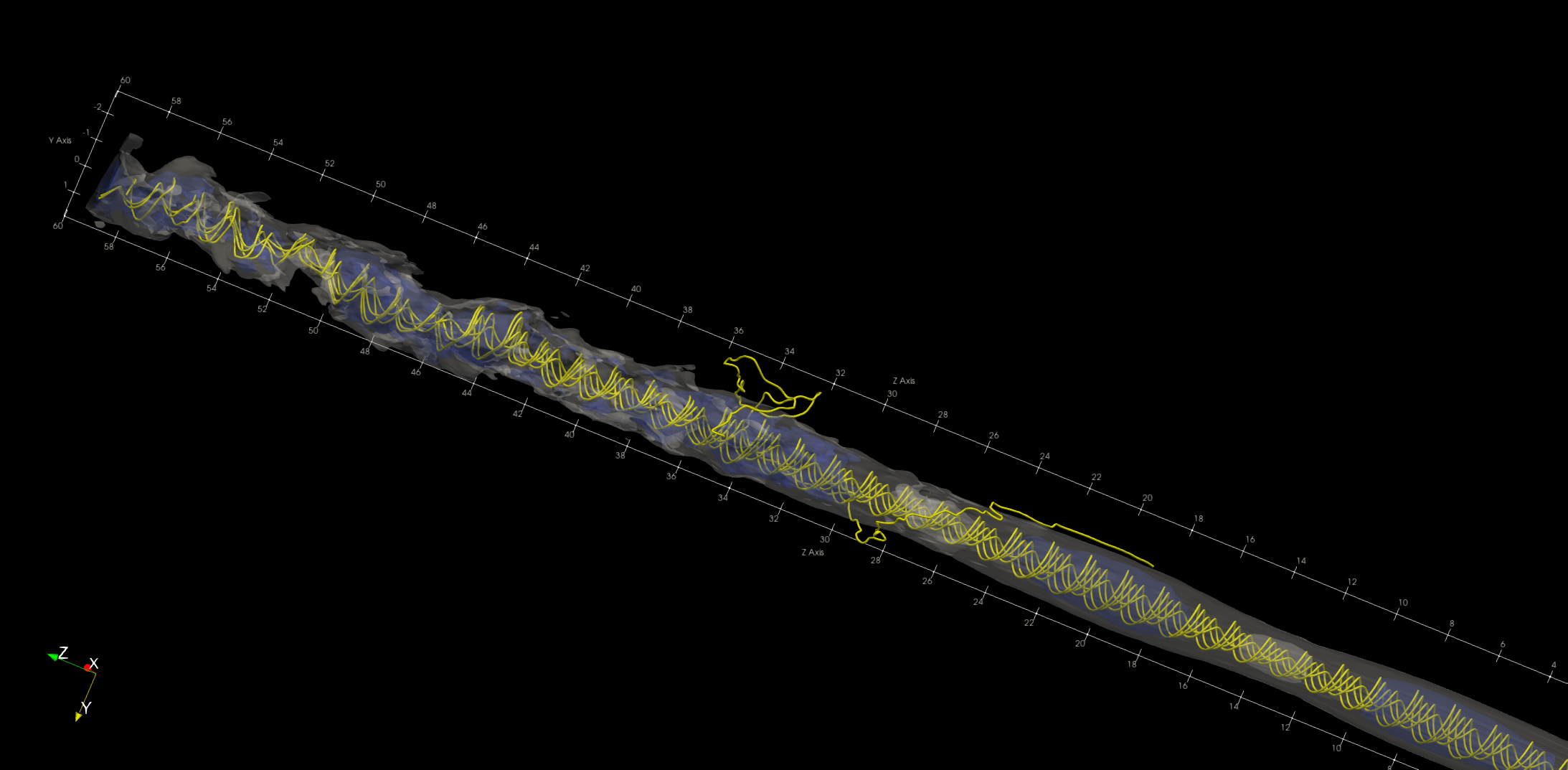}
    \caption{Three-dimensional density isosurface at $t=360$. The white(blue) surface marks the isosurface where density equals $0.01(0.005)$. The yellow lines are traces of the magnetic field lines. The color scales with the logarithm of the density. 
    }
    \label{fig:kink}
\end{figure*}

As seen in the effect of the magnetic field in over-pressured relativistic jets, the existence of a magnetic field helps stabilize the jets against RTI. In some cases, RTI has almost ceased to grow. However, in strongly magnetized relativistic jets with helical magnetic fields, CD kink instability potentially grows.
Figure~\ref{fig:kink} shows 3D volume rendering of density with magnetic field lines for the case {\tt MHD1} at $t_s=360$. At $z>30\,R_j$, a helically twisted jet structure is observed. A similar structure is also seen in the {\tt MHD3} case (the middle panel in Fig.~\ref{fig:k}). 
Helical magnetic fields also follow helically twisted density structures. Similar behavior is seen in previous 3D RMHD simulations of CD kink instability in relativistic jets \citep[e.g.,][]{mizuno2009,mizuno2012,mizuno2014,singh2016} although some of the magnetic field lines show zigzag trajectories due to the turbulence. 
After the passing of recollimation shock, toroidal magnetic becomes stronger, and the relativistic jet remains highly magnetized.  
It is a good condition for the growth of CD kink instability.
The development of CD kink instability after the recollimation shock is seen in 3D RMHD simulations of jet propagation in the stratified external medium \citep{tchekhovskoy2016, duran2017}.

\begin{figure}[!h]
    \centering
    \includegraphics[width=1.0\columnwidth]{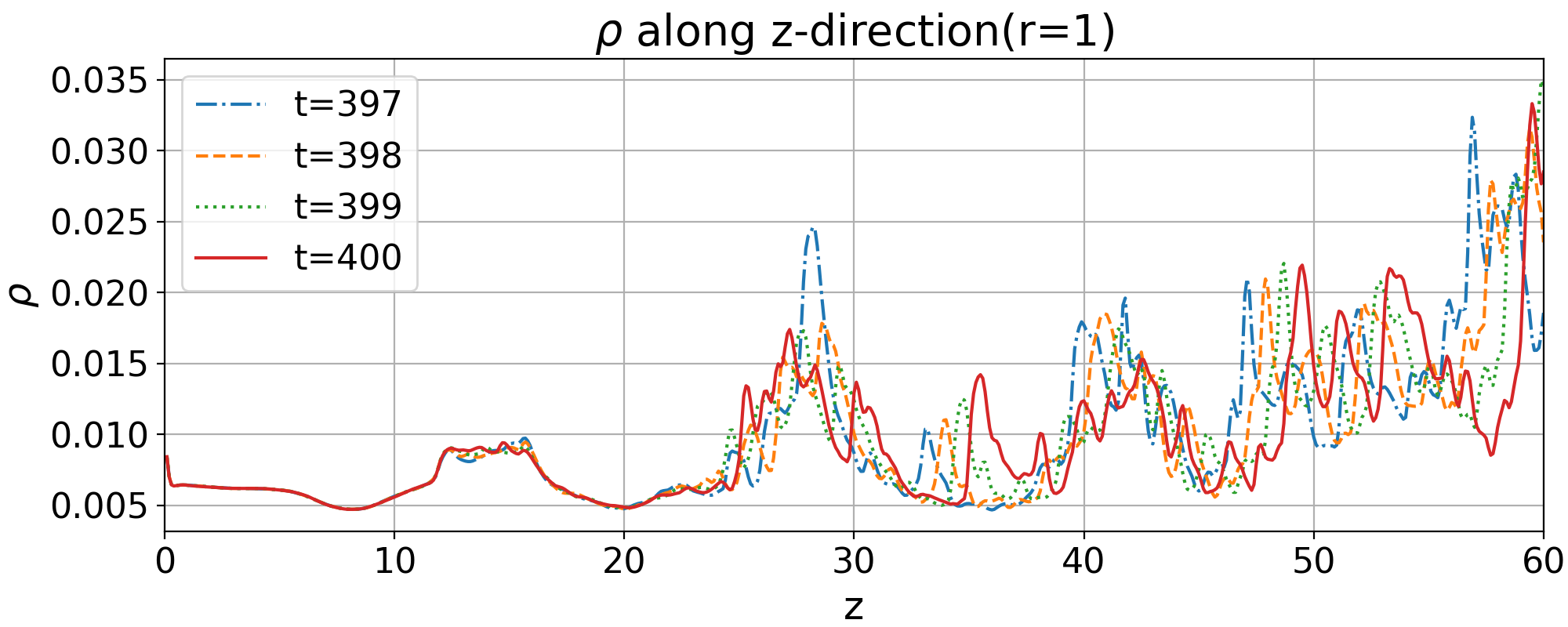}
    \caption{The azimuthal average density profiles at $R=1\,R_j$ for $t_s=397$ (blue dash-dotted), $398$ (orange dashed), $399$ (green dotted), and $400$ (red solid).
    }
    \label{fig:kink_speed}
\end{figure}

To investigate the propagation speed of the helically twisted structure, we plot the azimuthal-averaged density profiles at $R=1\,R_j$ in different times with $t_s=397$, $398$, $399$, and $400$ in Fig.\ref{fig:kink_speed}.

From this figure, we capture 4 peaks between $z=47\,R_j$ and $z=50\,R_j$. Thus, the propagation speed of a helically twisted structure is about $0.8c$, which is mostly consistent with the propagation speed of the jet itself. It is similar results seen in \citep[e.g.,][]{mizuno2014,singh2016}

\section{Summary and Discussion}\label{dis}
In this paper, we performed 3D RMHD simulations of over-pressured magnetized jets and explored the effect of the magnetic pitch and strength. We began our simulation with the 2D axisymmetric configuration and obtained a stable recollimation shock structure, retrieving the conclusion of \citet{mizuno2015}. From the simulations, we observed a low effective inertia interface between the jet and the ambient medium. 
By removing axisymmetric symmetry, the 3D RHD simulations of the over-pressured jets exhibited the development of RTI at the interface between jet and ambient medium, confirming the prediction of \cite{Matsumoto2017}.
For 3D RMHD simulations of over-pressured jets, we observed that the recollimation shock structure transforms into a mixture of the RTI at the interface between the jet and ambient medium and the CD kink instability in the jet.
As stated by \cite{millas2017}, increasing the toroidal magnetic field contribution (decreasing the magnetic pitch) stabilizes the jet against RTI. However,  our simulations did not obey the suggested stability criterion that the magnetization $\sigma>0.01$ keeps the jet stable. It implies that magnetic field configuration affects the jet stability criterion.

We derive the radial motion equation of the recollimation structure with the assumption that the radial magnetic field equals 0 and radial velocity is infinitesimal(Appendix.\ref{acce}). Pressure gradient, magnetic pressure, and centrifugal force affect radial acceleration. We confirmed that the radial motions of jets are consistent with the prediction of this equation both in 2D and 3D cases. Although we have not obtained the analytical expression of stability criteria in this work, these three kinds of force will all play their roles in the analysis of instability. It will be included in our future study.

This study is extension from our previous work \citep{mizuno2015}. Thus, we assume an external medium with simple contact density. However, in reality, the external medium would likely be characterized by declining profiles. Several papers have discussed the effect of declining density profiles on to the jet dynamics and structures \citep[e.g.,][]{gomez1997,porth2015,tchekhovskoy2016,gomez2016,duran2017,fromm2019}. In general, the jet has radially expanded and the distance between two recollimation shocks becomes larger with distance. 
We will discuss it in a separate paper.

In this work, we assume jet is relatively hot. However, in the nature, jet might be cold. A cold jet has a smaller effective inertia. It becomes lighter. Therefore, the deformation of the jet would be more obvious. However, if jet is over-pressured, similarly the recollimation shock is developed and RTI will destrupt the interface between jet and external medium. The existence of magnetic field would be stabilize the RTI as seen in this work.Thus, we expect the global picture may not change much.

In our 3D RMHD simulations with a helical magnetic field, the excitement of CD kink instability is seen after passing the recollimation shocks.
Recently \cite{Jortstad2022}  suggested the development of CD kink instability after passed recollimation shock to explain the quasi-periodic oscillations (QPOs) observed during the outburst in 2020 of blazar BL Lac. 
BL Lac has recognized the existence of multiple stationary knots \citep{cohen2014,gomez2016} that would be multiple recollimation shocks. \cite{cohen2015}  suggested the existence of a helical magnetic field in relativistic jets of BL Lac. \cite{Jortstad2022}  also argued that the growth of the kink with time could match the QSO properties in BL Lac. Our simulations are qualitatively consistent with their observations. 
We think our simulation results provide a consistent picture with the observations in BL Lac. 
However, in this work, our numerical setup is idealized for the investigations of instability.
For future studies, we will set up the initial conditions of simulation matched with the condition in BL Lac to obtain observables such as jet images, polarization, and light curves that are directly comparable to the observations.

\begin{acknowledgements}
  This research is supported by the National Key R\&D Program of China (2023YFE0101200), the National Natural Science Foundation of China (Grant No. 12273022), and the Shanghai Municipality orientation program of Basic Research for International Scientists (Grant No. 22JC1410600). 
  CMF is supported by the DFG research grant “Jet physics on horizon scales and beyond" (Grant No. 443220636) within the DFG research unit “Relativistic Jets in Active Galaxies" (FOR 5195).
  The simulations were performed on the Astro cluster at Tsung-Dao Lee Institute and the Siyuan-1 cluster at the High-Performance Computing Center at Shanghai Jiao Tong University.
This work has used NASA's Astrophysics Data System (ADS).    
\end{acknowledgements}

%
   \bibliographystyle{aa} 
   \bibliography{main} 

\begin{thebibliography}{39}
\expandafter\ifx\csname natexlab\endcsname\relax\def\natexlab#1{#1}\fi

\bibitem[{{Abolmasov} \& {Bromberg}(2023)}]{abolmasov2023}
{Abolmasov}, P. \& {Bromberg}, O. 2023, \mnras, 520, 3009

\bibitem[{{Agudo} {et~al.}(2001){Agudo}, {G{\'o}mez}, {Mart{\'\i}}, {Ib{\'a}{\~n}ez}, {Marscher}, {Alberdi}, {Aloy}, \& {Hardee}}]{agudo2001}
{Agudo}, I., {G{\'o}mez}, J.-L., {Mart{\'\i}}, J.-M., {et~al.} 2001, \apjl, 549, L183

\bibitem[{{Barniol Duran} {et~al.}(2017){Barniol Duran}, {Tchekhovskoy}, \& {Giannios}}]{duran2017}
{Barniol Duran}, R., {Tchekhovskoy}, A., \& {Giannios}, D. 2017, \mnras, 469, 4957

\bibitem[{Blandford \& Payne(1982)}]{blandford1982}
Blandford, R.~D. \& Payne, D.~G. 1982, Monthly Notices of the Royal Astronomical Society, 199, 883

\bibitem[{{Blandford} \& {Znajek}(1977)}]{blandford1977}
{Blandford}, R.~D. \& {Znajek}, R.~L. 1977, \mnras, 179, 433

\bibitem[{{Cheung} {et~al.}(2007){Cheung}, {Harris}, \& {Stawarz}}]{cheung2007}
{Cheung}, C.~C., {Harris}, D.~E., \& {Stawarz}, {\L}. 2007, \apjl, 663, L65

\bibitem[{{Cohen} {et~al.}(2015){Cohen}, {Meier}, {Arshakian}, {Clausen-Brown}, {Homan}, {Hovatta}, {Kovalev}, {Lister}, {Pushkarev}, {Richards}, \& {Savolainen}}]{cohen2015}
{Cohen}, M.~H., {Meier}, D.~L., {Arshakian}, T.~G., {et~al.} 2015, \apj, 803, 3

\bibitem[{{Cohen} {et~al.}(2014){Cohen}, {Meier}, {Arshakian}, {Homan}, {Hovatta}, {Kovalev}, {Lister}, {Pushkarev}, {Richards}, \& {Savolainen}}]{cohen2014}
{Cohen}, M.~H., {Meier}, D.~L., {Arshakian}, T.~G., {et~al.} 2014, \apj, 787, 151

\bibitem[{{Costa} {et~al.}(2023){Costa}, {Bodo}, {Tavecchio}, {Rossi}, {Capetti}, {Massaglia}, {Sciaccaluga}, {Baldi}, \& {Giovannini}}]{costa2023}
{Costa}, A., {Bodo}, G., {Tavecchio}, F., {et~al.} 2023, arXiv e-prints, arXiv:2312.08767

\bibitem[{{Curtis}(1918)}]{curtis1918}
{Curtis}, H.~D. 1918, Publications of Lick Observatory, 13, 55

\bibitem[{{Fanaroff} \& {Riley}(1974)}]{fanaroff1974}
{Fanaroff}, B.~L. \& {Riley}, J.~M. 1974, \mnras, 167, 31P

\bibitem[{{Fromm} {et~al.}(2016){Fromm}, {Perucho}, {Mimica}, \& {Ros}}]{fromm2016}
{Fromm}, C.~M., {Perucho}, M., {Mimica}, P., \& {Ros}, E. 2016, \aap, 588, A101

\bibitem[{{Fromm} {et~al.}(2019){Fromm}, {Younsi}, {Baczko}, {Mizuno}, {Porth}, {Perucho}, {Olivares}, {Nathanail}, {Angelakis}, {Ros}, {Zensus}, \& {Rezzolla}}]{fromm2019}
{Fromm}, C.~M., {Younsi}, Z., {Baczko}, A., {et~al.} 2019, \aap, 629, A4

\bibitem[{{Ghisellini}(2011)}]{ghisellini2011}
{Ghisellini}, G. 2011, in American Institute of Physics Conference Series, Vol. 1381, 25th Texas Symposium on Relativistic AstroPhysics (Texas 2010), ed. F.~A. {Aharonian}, W.~{Hofmann}, \& F.~M. {Rieger}, 180--198

\bibitem[{{Giroletti} {et~al.}(2012){Giroletti}, {Hada}, {Giovannini}, {Casadio}, {Beilicke}, {Cesarini}, {Cheung}, {Doi}, {Krawczynski}, {Kino}, {Lee}, \& {Nagai}}]{giroletti2012}
{Giroletti}, M., {Hada}, K., {Giovannini}, G., {et~al.} 2012, \aap, 538, L10

\bibitem[{{G{\'o}mez} {et~al.}(2016){G{\'o}mez}, {Lobanov}, {Bruni}, {Kovalev}, {Marscher}, {Jorstad}, {Mizuno}, {Bach}, {Sokolovsky}, {Anderson}, {Galindo}, {Kardashev}, \& {Lisakov}}]{gomez2016}
{G{\'o}mez}, J.~L., {Lobanov}, A.~P., {Bruni}, G., {et~al.} 2016, \apj, 817, 96

\bibitem[{{G{\'o}mez} {et~al.}(1997){G{\'o}mez}, {Mart{\'\i}}, {Marscher}, \& {Ib{\'a}{\~n}ez}}]{gomez1997}
{G{\'o}mez}, J.~L., {Mart{\'\i}}, J.~M., {Marscher}, A.~P., \& {Ib{\'a}{\~n}ez}, J.~M. 1997, Vistas in Astronomy, 41, 79

\bibitem[{{Gourgouliatos} \& {Komissarov}(2018)}]{Gourgouliatos2018}
{Gourgouliatos}, K.~N. \& {Komissarov}, S.~S. 2018, Nature Astronomy, 2, 167

\bibitem[{{Jorstad} {et~al.}(2005){Jorstad}, {Marscher}, {Lister}, {Stirling}, {Cawthorne}, {Gear}, {G{\'o}mez}, {Stevens}, {Smith}, {Forster}, \& {Robson}}]{jorstad2005}
{Jorstad}, S.~G., {Marscher}, A.~P., {Lister}, M.~L., {et~al.} 2005, \aj, 130, 1418

\bibitem[{{Jorstad} {et~al.}(2022){Jorstad}, {Marscher}, {Raiteri}, {Villata}, {Weaver}, {Zhang}, {Dong}, {G{\'o}mez}, {Perel}, {Savchenko}, {Larionov}, {Carosati}, {Chen}, {Kurtanidze}, {Marchini}, {Matsumoto}, {Mortari}, {Aceti}, {Acosta-Pulido}, {Andreeva}, {Apolonio}, {Arena}, {Arkharov}, {Bachev}, {Banfi}, {Bonnoli}, {Borman}, {Bozhilov}, {Carnerero}, {Damljanovic}, {Ehgamberdiev}, {Els{\"a}sser}, {Frasca}, {Gabellini}, {Grishina}, {Gupta}, {Hagen-Thorn}, {Hallum}, {Hart}, {Hasuda}, {Hemrich}, {Hsiao}, {Ibryamov}, {Irsmambetova}, {Ivanov}, {Joner}, {Kimeridze}, {Klimanov}, {Kn{\"o}tt}, {Kopatskaya}, {Kurtanidze}, {Kurtenkov}, {Kuutma}, {Larionova}, {Leonini}, {Lin}, {Lorey}, {Mannheim}, {Marino}, {Minev}, {Mirzaqulov}, {Morozova}, {Nikiforova}, {Nikolashvili}, {Ovcharov}, {Papini}, {Pursimo}, {Rahimov}, {Reinhart}, {Sakamoto}, {Salvaggio}, {Semkov}, {Shakhovskoy}, {Sigua}, {Steineke}, {Stojanovic}, {Strigachev}, {Troitskaya}, {Troitskiy}, {Tsai}, {Valcheva}, {Vasilyev}, {Vince}, {Waller}, {Zaharieva}, \&
  {Chatterjee}}]{Jortstad2022}
{Jorstad}, S.~G., {Marscher}, A.~P., {Raiteri}, C.~M., {et~al.} 2022, \nat, 609, 265

\bibitem[{Keppens \& Meliani(2012)}]{bottcher2012}
Keppens, R. \& Meliani, Z. 2012, Jet Structure, Collimation and Stability: Recent Results from Analytical Models and Simulations (John Wiley \& Sons, Ltd), 341--368

\bibitem[{{Komissarov} \& {Falle}(1997)}]{komissarov1997}
{Komissarov}, S.~S. \& {Falle}, S.~A.~E.~G. 1997, \mnras, 288, 833

\bibitem[{{Komissarov} {et~al.}(2019){Komissarov}, {Gourgouliatos}, \& {Matsumoto}}]{Komissarov2019}
{Komissarov}, S.~S., {Gourgouliatos}, K.~N., \& {Matsumoto}, J. 2019, \mnras, 488, 4061

\bibitem[{{Lister} {et~al.}(2013){Lister}, {Aller}, {Aller}, {Homan}, {Kellermann}, {Kovalev}, {Pushkarev}, {Richards}, {Ros}, \& {Savolainen}}]{lister2013}
{Lister}, M.~L., {Aller}, M.~F., {Aller}, H.~D., {et~al.} 2013, \aj, 146, 120

\bibitem[{{Matsumoto} {et~al.}(2017){Matsumoto}, {Aloy}, \& {Perucho}}]{Matsumoto2017}
{Matsumoto}, J., {Aloy}, M.~A., \& {Perucho}, M. 2017, \mnras, 472, 1421

\bibitem[{{Matsumoto} {et~al.}(2021){Matsumoto}, {Komissarov}, \& {Gourgouliatos}}]{matsumoto2021}
{Matsumoto}, J., {Komissarov}, S.~S., \& {Gourgouliatos}, K.~N. 2021, \mnras, 503, 4918

\bibitem[{{Matsumoto} \& {Masada}(2013)}]{Matsumoto2013}
{Matsumoto}, J. \& {Masada}, Y. 2013, \apjl, 772, L1

\bibitem[{{Meliani} \& {Keppens}(2009)}]{Meliani2009}
{Meliani}, Z. \& {Keppens}, R. 2009, \apj, 705, 1594

\bibitem[{{Mignone} {et~al.}(2007){Mignone}, {Bodo}, {Massaglia}, {Matsakos}, {Tesileanu}, {Zanni}, \& {Ferrari}}]{mignone2007}
{Mignone}, A., {Bodo}, G., {Massaglia}, S., {et~al.} 2007, \apjs, 170, 228

\bibitem[{{Millas} {et~al.}(2017){Millas}, {Keppens}, \& {Meliani}}]{millas2017}
{Millas}, D., {Keppens}, R., \& {Meliani}, Z. 2017, \mnras, 470, 592

\bibitem[{{Mizuno} {et~al.}(2015){Mizuno}, {G{\'o}mez}, {Nishikawa}, {Meli}, {Hardee}, \& {Rezzolla}}]{mizuno2015}
{Mizuno}, Y., {G{\'o}mez}, J.~L., {Nishikawa}, K.-I., {et~al.} 2015, \apj, 809, 38

\bibitem[{{Mizuno} {et~al.}(2014){Mizuno}, {Hardee}, \& {Nishikawa}}]{mizuno2014}
{Mizuno}, Y., {Hardee}, P.~E., \& {Nishikawa}, K.-I. 2014, \apj, 784, 167

\bibitem[{{Mizuno} {et~al.}(2009){Mizuno}, {Lyubarsky}, {Nishikawa}, \& {Hardee}}]{mizuno2009}
{Mizuno}, Y., {Lyubarsky}, Y., {Nishikawa}, K.-I., \& {Hardee}, P.~E. 2009, \apj, 700, 684

\bibitem[{{Mizuno} {et~al.}(2012){Mizuno}, {Lyubarsky}, {Nishikawa}, \& {Hardee}}]{mizuno2012}
{Mizuno}, Y., {Lyubarsky}, Y., {Nishikawa}, K.-I., \& {Hardee}, P.~E. 2012, \apj, 757, 16

\bibitem[{{Moll} {et~al.}(2008){Moll}, {Spruit}, \& {Obergaulinger}}]{moll2008}
{Moll}, R., {Spruit}, H.~C., \& {Obergaulinger}, M. 2008, \aap, 492, 621

\bibitem[{{Porth} \& {Komissarov}(2015)}]{porth2015}
{Porth}, O. \& {Komissarov}, S.~S. 2015, \mnras, 452, 1089

\bibitem[{{Pushkarev} {et~al.}(2009){Pushkarev}, {Kovalev}, {Lister}, \& {Savolainen}}]{pushkarev2009}
{Pushkarev}, A.~B., {Kovalev}, Y.~Y., {Lister}, M.~L., \& {Savolainen}, T. 2009, \aap, 507, L33

\bibitem[{{Singh} {et~al.}(2016){Singh}, {Mizuno}, \& {de Gouveia Dal Pino}}]{singh2016}
{Singh}, C.~B., {Mizuno}, Y., \& {de Gouveia Dal Pino}, E.~M. 2016, \apj, 824, 48

\bibitem[{{Tchekhovskoy} \& {Bromberg}(2016)}]{tchekhovskoy2016}
{Tchekhovskoy}, A. \& {Bromberg}, O. 2016, \mnras, 461, L46

\end{thebibliography}

\appendix
\section{Acceleration in recollimation structure} \label{acce}

Although instabilities destroy the structures of 3D jets, recollimation shocks can be still observed near the jet inlet ($z=0$). \cite{Matsumoto2017}  considers the radial acceleration of the recollimation shocks as the effective gravity causing the RTI. Here we give an analysis for a more general condition.

At first, we assume there is a stable axisymmetric jet with $v_R\ll v_z$ and there is no radial component of the magnetic field $B_R=0$. The first term of Equation~(\ref{2}) can be expanded and simplified as :
\begin{equation}
    (\gamma^2\rho h+B_\phi^2+B_z^2)v_R=I v_R, \label{d1}
\end{equation}
where $I$ is effective inertia. It shows both toroidal and poloidal magnetic fields serve as a part of mass somehow. 
The second term of Eq.~(\ref{2}) can be expressed as:
\begin{equation}  
\begin{aligned}
&[\nabla\cdot(\omega_t\gamma^2\mathbf{v}\mathbf{v}-\mathbf{b}\mathbf{b}+\mathbf{I} p_t)]\cdot\hat r \\
&=[(\nabla\cdot(\gamma^2\omega_t \mathbf{v}-b^0\mathbf{b}))\mathbf{v}+(\gamma^2\omega_t \mathbf{v}\cdot\nabla)\mathbf{v} \\
&\quad\ +(\nabla b^0\cdot\mathbf{b})\mathbf{v}-(\mathbf{b}\cdot \nabla)\mathbf{b}]\cdot\hat r +\frac{\partial p_t}{\partial R}. \label{d2}
    \end{aligned}
  \end{equation}
Thus, the radial component of the momentum equation can be written as:
\begin{equation}
    \begin{aligned}        
    &\frac{\partial}{\partial t}(I v_R)+[(\nabla\cdot(\gamma^2\omega_t \mathbf{v}-b^0\mathbf{b}))\mathbf{v}+(\gamma^2\omega_t \mathbf{v}\cdot\nabla)\mathbf{v}\\
    &+(\nabla b^0\cdot\mathbf{b})\mathbf{v}-(\mathbf{b}\cdot \nabla)\mathbf{b}]\cdot\hat r+\frac{\partial p_t}{\partial R}=0\label{d3}
    \end{aligned}
\end{equation}
Notice that the first term in Equation~(\ref{3}) can be simplified as $\gamma^2\rho h+B_\phi^2+B_z^2-p_t$. Then we multiply Equation~(\ref{3}) with $v_R$:
\begin{equation}
    \begin{aligned}
       v_R\frac{\partial }{\partial t}(I-p_t)+[\nabla\cdot(\omega_t\gamma^2\mathbf{v}-b^0\mathbf{b})]v_R&=0\\
       \frac{\partial }{\partial t}(I v_R)-I\frac{\partial v_R}{\partial t}-v_R\frac{\partial p_t}{\partial t}+[\nabla\cdot(\omega_t\gamma^2\mathbf{v}-b^0\mathbf{b})]v_R&=0\\
    \end{aligned}\label{d4}
\end{equation}
Here we can ignore the third and fourth terms since $v_R\sim0$. Finally, we subtract equation~(\ref{d3}) from equation~(\ref{d4}):
\begin{equation}
    I\frac{\partial v_R}{\partial t}+\frac{\partial p_t}{\partial R}+[(\gamma^2\omega_t \mathbf{v}\cdot\nabla)\mathbf{v}+(\nabla b^0\cdot\mathbf{b})\mathbf{v}-(\mathbf{b}\cdot \nabla)\mathbf{b}]\cdot\hat r=0 \label{d5}
\end{equation}
By expanding all terms, it is written as:
\begin{equation}
    \begin{aligned}
    [(\gamma^2\omega_t \mathbf{v}\cdot\nabla)\mathbf{v}]\cdot\hat{r}&=-\frac{\gamma^2\omega_t v_\phi^2}{R}\\
    [(\nabla b^0\cdot\mathbf{b})\mathbf{v}]\cdot\hat{r}&=0\\
    [-(\mathbf{b}\cdot \nabla)\mathbf{b}]\cdot\hat{r}&=\frac{b_\phi^2}{R}
    \end{aligned}
\end{equation}

\begin{figure}[!ht]
    \centering
    \includegraphics[width=0.45\textwidth]{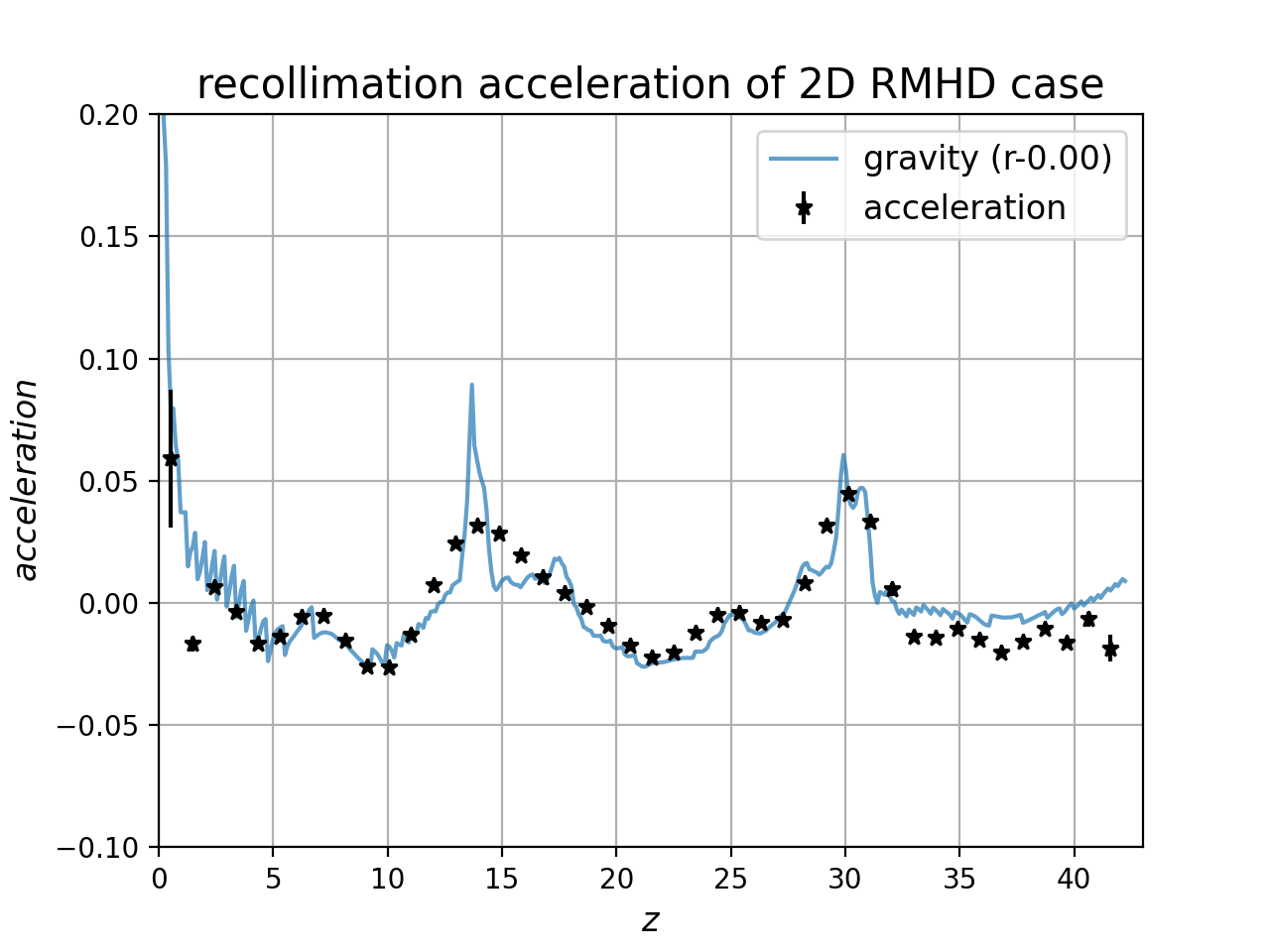}
    \includegraphics[width=0.45\textwidth]{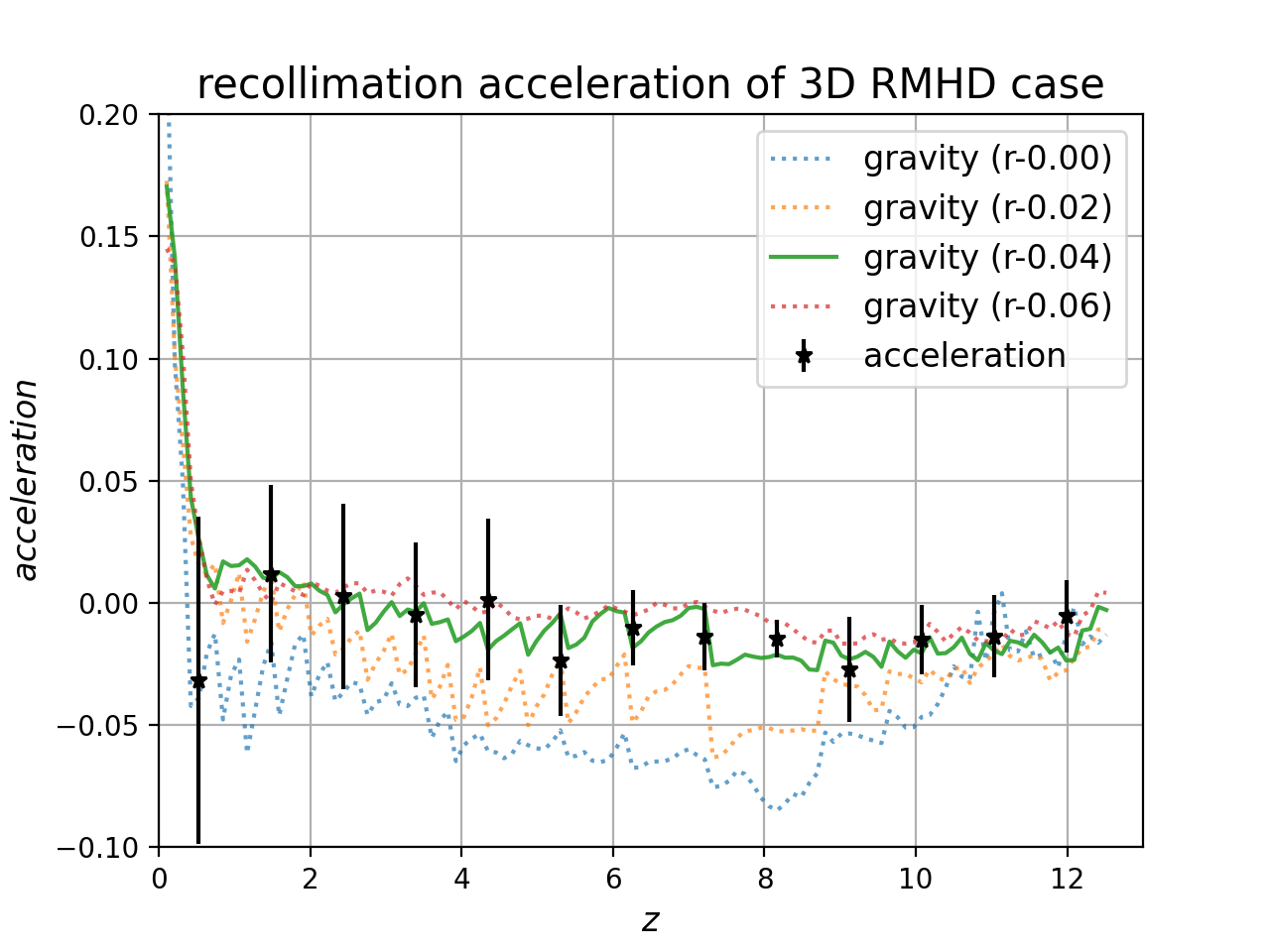}
    \caption{Axial distribution of the acceleration by effective gravity at the jet boundary for 2D RMHD case {\tt MHD1-2D} ({\it top}) and 3D RMHD case {\tt MHD1} ({\it bottom}). Different color lines indicate the effective gravity at different radii of simulations, $R=R_{jb}$ (blue dotted), $R_{jb}-0.02$ (orange dotted), $R_{jb}-0.04$ (green solid), and $R_{jb}-0.06$, where $R_{jb}$ is the radius of the jet boundary. Star marks are the azimuthal averaged radial acceleration calculated directly from the radius variations. The error bar is the standard deviation of the averages.}
    \label{fig:re}
\end{figure}

%
Here we use the condition $v_R\sim0$ again. Now we obtain the equation describing the radial motion of the recollimated jet:
\begin{equation}
    I\frac{\partial v_R}{\partial t}+\frac{\partial p_t}{\partial R}-\frac{\gamma^2\omega_t v_\phi^2}{R}+\frac{b_\phi^2}{R}=0\label{d6}
\end{equation}

To compare this formula with the simulations, we calculate the acceleration in two approaches: the kinematic method which calculates the changes of the jet radius directly, and the dynamic method which calculates the effective gravity. To determine the radius of the jet boundary $R_{jb}$, we follow \cite{Meliani2009}, counting the number of grids that have high Lorentz factors. Then we calculate the effective gravity through 
\begin{equation}
g=\left(-\frac{\partial p_t}{\partial R}+\frac{\gamma^2\omega_t v_\phi^2}{R}-\frac{b_\phi^2}{R}\right)/I.
\end{equation}
Because of the existence of turbulence at the jet boundary, we do not apply the gravity value exactly at the interface. To avoid turbulence, we use a little smaller radius from the jet boundary. This will not affect the conclusion because gravity inside the jet varies slowly with the radius. The turbulence also causes a large dispersion of the acceleration derived from the kinematic method. Thus, we take the average acceleration of every 9 neighboring grids as the final result.

Figure~\ref{fig:re} shows the comparison of accelerations from two approaches. The top panel of Fig.~\ref{fig:re} presents a comparison in the 2D RMHD case.
We use a polynomial fit to the radius as a function of time. Without the disturbance of the RTI, the acceleration matches the theoretical calculations perfectly except around the narrow peak. Because the polynomial does not fit the radius well around the peak. 
The bottom  panel of Figure~\ref{fig:re} shows the comparison for the 3D RMHD case {\tt MHD1}. The effective gravity at $R_{jb}-0.04$ is nearly the same as the effective gravity at $R_{jb}-0.06$, implying the motion of the jet is not perturbed by RTI at this radius. 
Thus, we found it is quite comparable with analytical calculations although it has a relatively large dispersion.

\section{Effect of the boundary condition} \label{boundary}

\begin{figure*}[!ht]
    \centering
    \subfigure[]{\includegraphics[width=0.20\textwidth]{figure/3D_HD_rho.png}}
    \subfigure[]{\includegraphics[width=0.20\textwidth]{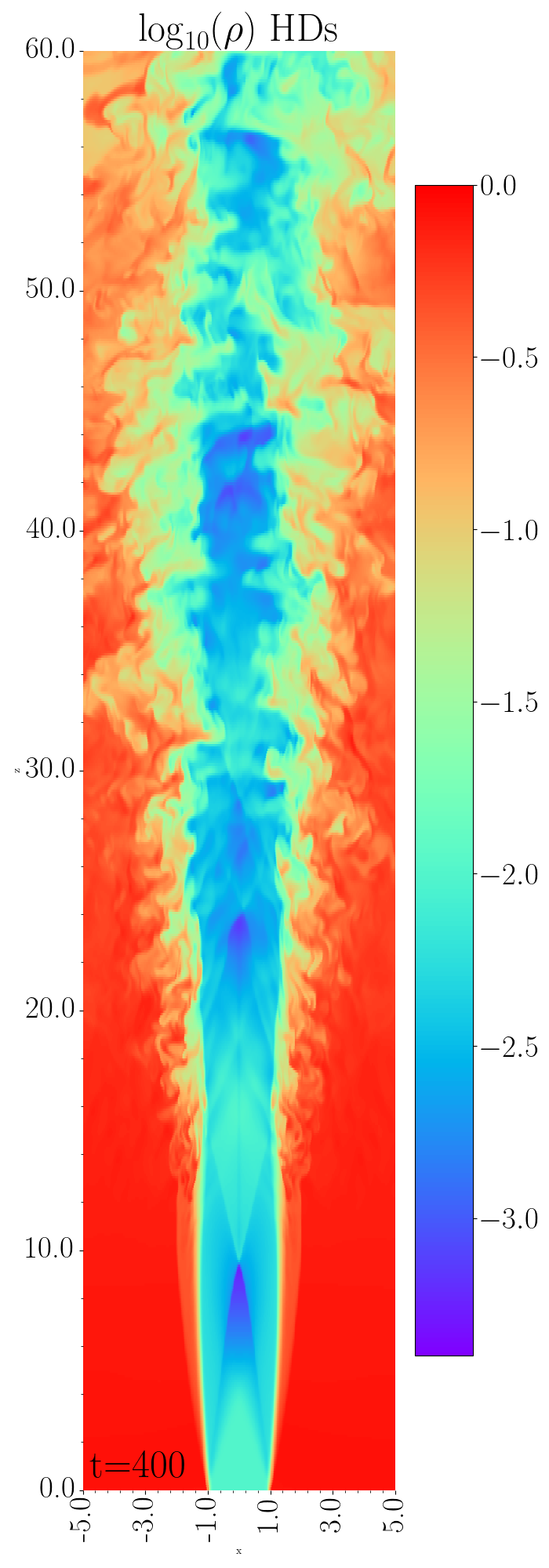}}
    \subfigure[]{\includegraphics[width=0.20\textwidth]{figure/3D_rho.png}}
    \subfigure[]{\includegraphics[width=0.20\textwidth]{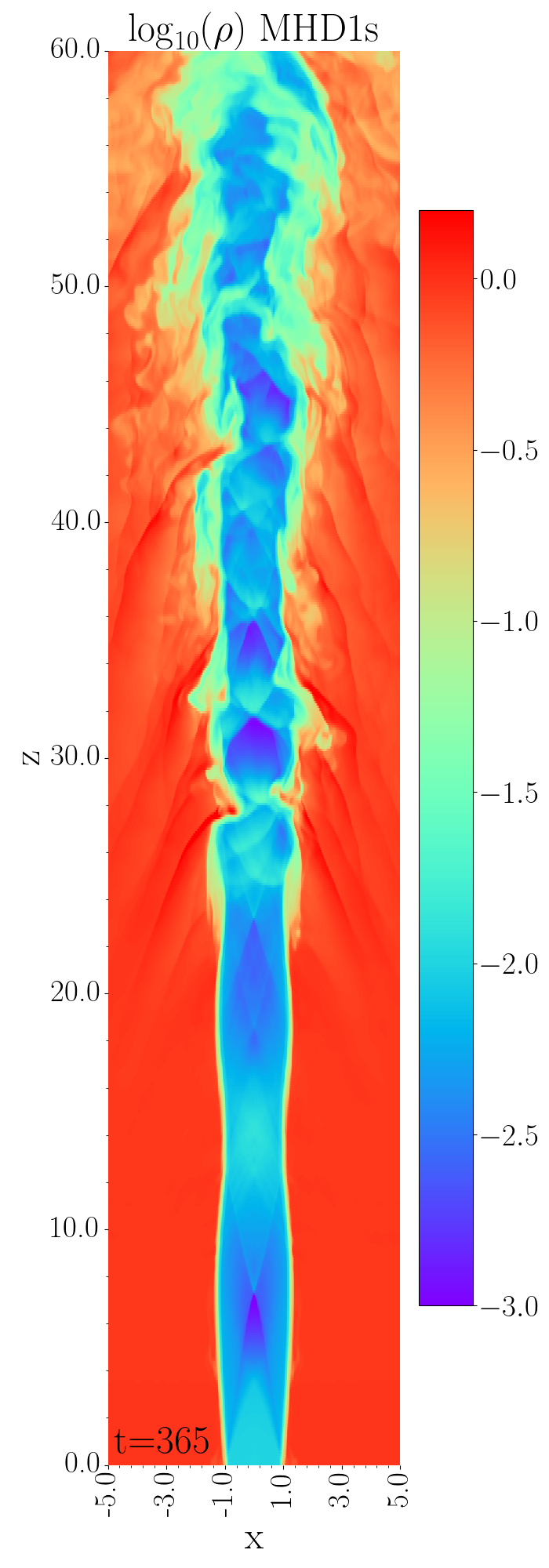}}
    \caption{2D axial distributions of logarithmic density for sharp boundary ({\it a, c}) and smooth boundary ({\it b, d}) for {\tt HD} case (left two panels) at $t_s=400$ and {\tt MHD1} case (right two panels) at $t_s=365$.}
    \label{fig:smooth}
\end{figure*}

Since the jet in our simulation is over-pressured than the external medium, not in equilibrium, we apply a sharp boundary between jet and external medium to make it expand freely when simulation starts. However, as discussed by e.g. \cite{abolmasov2023}, such sharp boundary would produce numerical noise and thus introduce spurious effects. In order to test it, we perform simulations with a smooth boundary through the approach developed by \cite{abolmasov2023}. Here, we choose $\Delta R=0.04R_j$ so that we can achieve a smooth boundary transition in $0.2R_j$.

Figure~\ref{fig:smooth} shows 2D axial distribution of logarithmic density for {\tt HD} and {\tt MHD1} cases with two different boundary conditions, sharp boundary (a,c) and smooth boundary (b,d).
In hydrodynamic case (left two panels), these two different boundary scenarios are mostly similar except that the size of the recollimation structure. In the smooth boundary case, it is a little smaller than that with a sharp boundary. However, RTI is triggered in both cases. 
For the {\tt MHD1} case (right two panels), the conclusion is the same. Smooth boundary relieves the disruption of jet but RTI and CD kink instability still evolve.

In summary, boundary conditions do not affect the main results qualitatively.

\section{A higher jet Lorentz factor} \label{lor}

\begin{figure}[!h]
    \centering
    \includegraphics[width=0.20\textwidth]{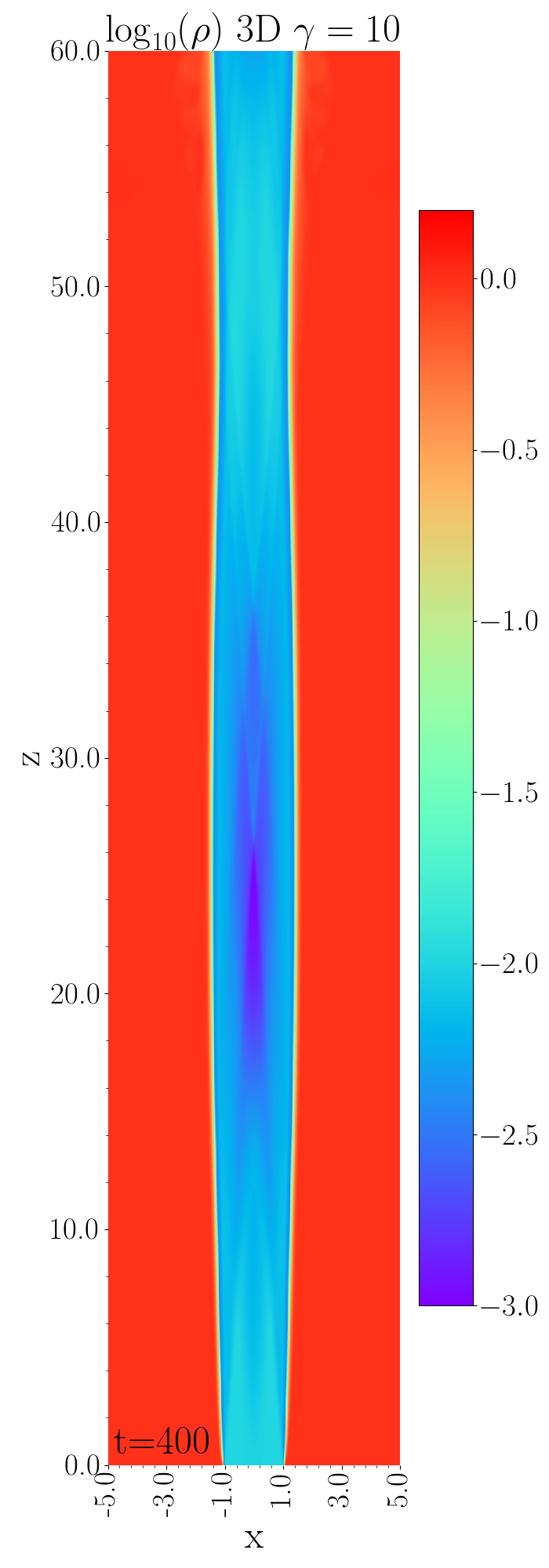}
    \caption{Axial distribution of logarithmic density for {\tt MHD1} with higher jet Lorentz factor, $\gamma=10$ at $t_s = 400$.}
    \label{fig:lor}
\end{figure}

The bulk Lorentz factor in this work is fixed to a moderate value of 3. Here we explore the impact of a larger Lorentz factor of jets with $\gamma_j=10$ in the {\tt MHD1} case, as commonly adopted for blazars. Due to increase of jet Lorentz factor, the effective inertia of the jet becomes 11 times higher than that in the original {\tt MHD1} case. We expect recollimation to be slower and on a smaller scale.

As is shown in Figure~\ref{fig:lor} shows 2D axial distribution of logarithmic density of higher jet Lorentz factor case. As we expected, large effective inertia slows down the growth of instability, and therefore the jet propagate for a longer distance stability that is consistent with the simulation of \cite{tchekhovskoy2016}.

For a rough estimation, we treat the recollimation approximation as an oscillation. We obtain the length of a recollimation structure $l\propto \sqrt{1/I}\propto1/\gamma$. The length in the {\tt MHD1} case is around $14$ and in the higher jet Lorentz factor case is $49$. We confirm that $49/14\simeq10/3$ is ratio of the initial jet Lorentz factor.

\end{document}